\documentclass[longauth]{aa} 

\usepackage{subcaption}
\usepackage{graphicx}
\usepackage{txfonts}
\usepackage[colorlinks=true,citecolor=blue]{hyperref}
\usepackage[normalem]{ulem}

\DeclareUnicodeCharacter{2212}{-}

\def\het{He(2$^{3}$S)}
\def\mlr{$\dot M$}
\def\lya{Ly$\alpha$}
\def\ha{H$\alpha$}

\def\hdu18{HD\,189733\,b}
\def\hdj20{HD\,209458\,b}
\def\gj34{GJ\,3470\,b}
\def\gju12{GJ\,1214\,b}
\def\w69{WASP-69\,b}
\def\wa76{WASP-76\,b}
\def\hatp32{HAT-P-32\,b}
\def\gs{g\,s$^{-1}$}

\def\rp{$R_{\rm P}$}

\def\hd23{HD\,235088}

\newcommand{\tachar}[1]{\bgroup\markoverwith{\textcolor{red}{\rule[0.5ex]{2pt}{1pt}}}\ULon{#1}}
\newcommand{\change}[2]{\bgroup\markoverwith{\textcolor{red}{\rule[0.5ex]{2pt}{1pt}}}\ULon{#1} \textcolor{blue} {#2}}

\begin{document}

   \title{Confirmation of an \ion{He}{I} evaporating atmosphere around the 650-Myr-old sub-Neptune HD\,235088\,b (TOI-1430\,b) with CARMENES}

   \titlerunning{Confirmation of \ion{He}{I} in HD\,235088\,b's atmosphere}

   \author{
J.~Orell-Miquel\inst{\ref{ins:iac},\ref{ins:ull}} \and
M.~Lamp\'on\inst{\ref{ins:iaa}} \and 
M.~L\'opez-Puertas\inst{\ref{ins:iaa}} \and 
M.~Mallorqu\'in\inst{\ref{ins:iac},\ref{ins:ull}} \and 
F.~Murgas\inst{\ref{ins:iac},\ref{ins:ull}}\and 
A.~Pel\'aez-Torres \inst{\ref{ins:iac},\ref{ins:ull}} \and 
E.~Pall\'e\inst{\ref{ins:iac},\ref{ins:ull}}\and 
E.~Esparza-Borges \inst{\ref{ins:iac},\ref{ins:ull}} \and 
J.~Sanz-Forcada\inst{\ref{ins:madrid}} \and 
H.~M.~Tabernero\inst{\ref{ins:CAB2}} \and
L.~Nortmann\inst{\ref{ins:eva1}} \and 
E.~Nagel\inst{\ref{ins:eva1}} \and 
H.~Parviainen\inst{\ref{ins:iac},\ref{ins:ull}} \and 
%
%
M.\,R.~Zapatero~Osorio\inst{\ref{ins:CAB2}}\and 
J.\,A.~Caballero\inst{\ref{ins:CAB2}} \and
S.~Czesla\inst{\ref{ins:eva2}} \and 
C.~Cifuentes\inst{\ref{ins:CAB2}} \and 
G.~Morello\inst{\ref{ins:iac},\ref{ins:sweden}}\and 
A.~Quirrenbach\inst{\ref{ins:LSW}} \and 
%
P.\,J.~Amado\inst{\ref{ins:iaa}} \and
A.~Fern\'andez-Mart\'in\inst{\ref{ins:CAHA}} \and 
A.~Fukui\inst{\ref{ins:astrojapan},\ref{ins:iac}} \and 
Th.~Henning\inst{\ref{ins:MPIA}} \and 
K.~Kawauchi\inst{\ref{ins:kiyoe}} \and 
J.\,P.~de Leon\inst{\ref{ins:jerome}} \and 
K.~Molaverdikhani\inst{\ref{ins:LMU},\ref{ins:ORIGINS},\ref{ins:LSW}} \and 
D.~Montes\inst{\ref{ins:MADRID}} \and 
N.~Narita\inst{\ref{ins:Komba},\ref{ins:astrojapan},\ref{ins:iac}} \and 
A.~Reiners\inst{\ref{ins:eva1}} \and 
I.~Ribas\inst{\ref{ins:ICE},\ref{ins:IEEC}} \and
A.~S\'anchez-L\'opez\inst{\ref{ins:iaa}} \and 
A.~Schweitzer\inst{\ref{ins:Hamburg}} \and 
M.~Stangret\inst{\ref{ins:inaf}} 
F.~Yan\inst{\ref{ins:china}}
          }

   \institute{
        \label{ins:iac}Instituto de Astrofísica de Canarias (IAC), C/ V\'ia L\'actea s/n, E-38205 La Laguna, Tenerife, Spain\\
        \email{jom@iac.es}
        \and
        \label{ins:ull}Departamento de Astrofísica, Universidad de La Laguna (ULL), Avd. Astrof\'isico Francisco S\'anchez s/n, E-38206 La Laguna, Tenerife, Spain
        \and
        \label{ins:iaa}Instituto de Astrof{\'i}sica de Andaluc{\'i}a (IAA-CSIC), Glorieta de la Astronom{\'i}a s/n, E-18008 Granada, Spain
        \and
        \label{ins:madrid}Centro de Astrobiología (CSIC-INTA), ESAC, Camino Bajo del Castillo s/n, Villanueva de la Cañada, E-28692 Madrid, Spain
        \and
        \label{ins:CAB2}Centro de Astrobiolog\'ia (CSIC-INTA), Carretera de Ajalvir km 4, E-28850 Torrej\'on de Ardoz, Madrid, Spain
        \and
        \label{ins:eva1}Institut f\"ur Astrophysik und Geophysik, Georg-August-Universit\"at, Friedrich-Hund-Platz 1, 37077 G\"ottingen, Germany
        \and
        \label{ins:eva2}Th\"uringer Landessternwarte Tautenburg, Sternwarte 5, 07778 Tautenburg, Germany
        \and
        \label{ins:sweden}Department of Space, Earth and Environment, Chalmers University of Technology, SE-412 96 Gothenburg, Sweden
        \and
        \label{ins:LSW}Landessternwarte, Zentrum f\"ur Astronomie der Universit\"at Heidelberg, K\"onigstuhl 12, D-69117 Heidelberg, Germany
        \and
        \label{ins:CAHA}Centro Astron\'omico Hispano en Andaluc\'\i{}a, Observatorio de Calar Alto, Sierra de los Filabres, E-04550 G\'ergal, Almer\'\i{}a, Spain
        \and
        \label{ins:astrojapan}Astrobiology Center, 2-21-1 Osawa, Mitaka, Tokyo 181-8588, Japan
        \and
        \label{ins:MPIA}Max-Planck-Institute f\"ur Astronomie, K\"onigstuhl 17, D-69117 Heidelberg, Germany
        \and
        \label{ins:kiyoe}Department of Multi-Disciplinary Sciences, Graduate School of Arts and Sciences, The University of Tokyo, 3-8-1 Komaba, Meguro, Tokyo 153-8902, Japan
        \and
        \label{ins:jerome}Department of Astronomy, Graduate School of Science, The University of Tokyo, 7-3-1 Hongo, Bunkyo-ku, Tokyo 113-0033, Japan
        \and
        \label{ins:LMU}Universit\"ats-Sternwarte, Ludwig-Maximilians-Universit\"at M\"unchen, Scheinerstrasse 1, D-81679 M\"unchen, Germany
        \and
        \label{ins:ORIGINS}Exzellenzcluster Origins, Boltzmannstrasse 2, 85748 Garching, Germany
        \and
        \label{ins:MADRID}Departamento de F{\'i}sica de la Tierra y Astrof{\'i}sica \& IPARCOS-UCM (Instituto de F\'{i}sica de Part\'{i}culas y del Cosmos de la UCM), Facultad de Ciencias F{\'i}sicas, Universidad Complutense de Madrid, E-28040 Madrid, Spain
        \and
        \label{ins:Komba}Komaba Institute for Science, The University of Tokyo, 3-8-1 Komaba, Meguro, Tokyo 153-8902, Japan
        \and
        \label{ins:ICE}Institut de Ci\`encies de l'Espai (ICE, CSIC), Campus UAB, Can Magrans s/n, 08193 Bellaterra, Barcelona, Spain
        \and
        \label{ins:IEEC}Institut d’Estudis Espacials de Catalunya (IEEC), 08034 Barcelona, Spain
        \and
        \label{ins:Hamburg}Hamburger Sternwarte, Gojenbergsweg 112, 21029 Hamburg, Germany
        \and
        \label{ins:inaf}Osservatorio Astronomico di Padova, Vicolo dell’Osservatorio 5, 35122, Padova, Italy
        \and
        \label{ins:china}Department of Astronomy, University of Science and Technology of China, Hefei 230026, China
        }

   \date{Received 17 March 2023 / Accepted 20 June 2023}

 
  \abstract
  {\hd23 (TOI-1430) is a young star known to host a sub-Neptune-sized planet candidate. We validated the planetary nature of \hd23\,b with multiband photometry, refined its planetary parameters, and obtained a new age estimate of the host star, placing it at 600-800\,Myr. Previous spectroscopic observations of a single transit detected an excess absorption of \ion{He}{I} coincident in time with the planet candidate transit. Here, we confirm the presence of \ion{He}{I} in the atmosphere of \hd23\,b  with one transit observed with CARMENES. We also detected hints of variability in the strength of the helium signal, with an absorption of $-$0.91$\pm$0.11\,\%, which is slightly deeper (2$\sigma$) than the previous measurement.
  Furthermore, we simulated the \ion{He}{I} signal with a spherically symmetric 1D hydrodynamic model, finding that the upper atmosphere of \hd23\,b escapes hydrodynamically with a significant mass loss rate of (1.5--5)\,$\times$10$^{10}$\,\gs\, in a relatively cold outflow, with $T$\,=\,3125\,$\pm$375\,K, in the photon-limited escape regime.
  \hd23\,b ($R_{\rm p}$\,=\,2.045$\pm$0.075\,R$_{\oplus}$) is the smallest planet found to date with a solid atmospheric detection -- not just of \ion{He}{I} but any other atom or molecule. This positions it a benchmark planet for further analyses of evolving young sub-Neptune atmospheres.
  } 


   \maketitle
%

\section{Introduction}

Along with their stars, planets evolve and change over time. During their early stages of formation, they suffer severe changes in their physical and orbital properties due to internal and external forces (\citealp{Baruteau_2106}). Our knowledge of the predominance and timescales of these changes is limited and a greater number of atmospherically characterized planets is needed to constrain formation and evolution models (\citealp{Dawson_Johnson_2018, Owen_Lai_2018, Tinetti_2018}). In this context, the study of planets at early stages of evolution is crucial for a better comprehension of different processes, such as planet formation and migration, inflation and evaporation of the primary atmospheres of rocky-core planets (\citealp{Owen_Wu_2017}), and the formation of the ``radius gap'' in the radius distribution of small planets (1--4\,$R_{\oplus}$; \citealp{Fulton_2017, Fulton_2018}).

Stars that host young planets are particularly useful because some of their planets are predicted to have not yet lost their extended H/He-rich primordial atmospheres. The space missions \textit{Kepler} (\citealp{Kepler_telescope}), \textit{K2} (\citealp{K2_Mission}), and \textit{Transiting Exoplanet Survey Satellite} (\textit{TESS}; \citealp{TESS_Ricker}) have discovered several transiting young planets orbiting $<$1\,Gyr-old stars, such as K2-33\,b (\citealp{David_2016_K2-33b}), as well as the V1298\,Tau (\citealp{David_2019_V1298Tau}) and AU\,Mic systems (\citealp{Plavchan_2020_AUMic}). These young planets are interesting targets to study, but they are also extremely challenging to analyze due to the intrinsic stellar variations of their young host stars (\citealp{Palle_2020_AUMicb, Benatti_2021_DSTuc}).

Moreover, the study of possible evaporating atmospheres of the young mini Neptune population is of special interest in helping identify the origin of the ``radius gap.'' That is the case for the HD\,63433 system (\citealp{HD63433_Zhang}), HD\,73583\,b (TOI-560\,b, \citealp{Zhang_TOI560b}), TOI-1683.01, \hd23.01 (TOI-1430.01), and TOI-2076\,b (\citealp{Zhang_TOI560b, Zhang_young_planets}). Two observations with Keck/NIRSPEC spectrograph robustly confirmed the presence of \ion{He}{I} in the atmosphere of HD\,73583\,b. The \ion{He}{I} detections for TOI-1683.01, \hd23.01, and TOI-2076\,b were made with single-transit observations. The same partial transit of TOI-2076\,b was also observed from the same mountain with the InfraRed Doppler (IRD) spectrograph by \cite{TOI-1807_TOI-2076_Gaidos2022}. Both analyses got a similar result of an \ion{He}{I} excess absorption of $\sim$1\,$\%$ that persisted for a short time after the egress. However, they differ in terms of the interpretation of the feature: \citet{Zhang_young_planets} claimed the signal to be a planetary absorption, while \cite{TOI-1807_TOI-2076_Gaidos2022} attributed it to stellar variability.

In this work, we derive stellar parameters for \hd23 using a high signal-to-noise (S/N) spectrum from the CARMENES spectrograph. We obtain a new age estimation for this young K-type star and validate HD\,235088\,b (TOI-1430\,b) as a planet using multi-color photometry. With CARMENES high-resolution spectra, we confirm the previous detection of \ion{He}{I} and, by analysing this \het\ signal, we study the hydrodynamical escape of this planet and derive the temperature and mass-loss rate of its upper atmosphere.

\section{Observations and data analysis}

\begin{table}
\caption{\label{table - System Param} Stellar parameters of \hd23.
}
\centering
\resizebox{\columnwidth}{!}{%
\begin{tabular}{lcl}
\hline
\hline
\noalign{\smallskip}
Parameter & Value & Reference \\
\noalign{\smallskip}
\hline
\noalign{\smallskip}

Name & HD\,235088 & HD \\
 & TIC 293954617 & \textit{TESS}\\
 & TOI-1430 & TOI \\
 & HIP 98668 & HIP \\
\noalign{\smallskip} \hline

\noalign{\smallskip} 
\multicolumn{3}{c}{ \textit{Coordinates and spectral types}} \\
\noalign{\smallskip}

$\alpha$\,(J2000) & 20$^\mathrm{h}$\,02$^\mathrm{m}$\,27$^\mathrm{s}\!$.4 & \textit{Gaia} EDR3 \\
$\delta$\,(J2000) & $+$53º\,22$'$\,36$"\!$.5 & \textit{Gaia} EDR3 \\
Spectral type & K2\,V & Sect.\,\ref{subsect: Stellar parameters} \\

\noalign{\smallskip} 
\multicolumn{3}{c}{ \textit{Parallax and kinematics}} \\
\noalign{\smallskip}

$\pi$ ~[mas] & 24.25\,$\pm$\,0.01 & \textit{Gaia} EDR3\\
$d$ ~[pc] & $41.24\pm 0.02$ & \textit{Gaia} EDR3 \\
$\mu_{\alpha} \cos{\delta}$ ~[mas\,yr$^{-1}$] & 165.05\,$\pm$\,0.02 & \textit{Gaia} EDR3\\
$\mu_{\delta}$ ~[mas\,yr$^{-1}$] & 145.17\,$\pm$\,0.02 & \textit{Gaia} EDR3\\
$\gamma$\,$^{(a)}$ ~[km\,s$^{-1}$] & $-$27.370$\pm$\,0.002 & \textit{Gaia} DR2 \\
$U$ ~[km\,s$^{-1}$] & $-$41.75\,$\pm$\,0.02 & This work\\
$V$ ~[km\,s$^{-1}$] & $-$22.16\,$\pm$\,0.01 & This work\\
$W$ ~[km\,s$^{-1}$] & $-$19.03\,$\pm$\,0.02 & This work\\
RUWE & $0.966$ & \textit{Gaia} DR3\\

\noalign{\smallskip} 
\multicolumn{3}{c}{ \textit{Magnitudes}} \\
\noalign{\smallskip} 

$B$ ~[mag] & 10.129 $\pm$ 0.038 & TYC \\
$V$ ~[mag] &  9.19 $\pm$ 0.03  & HIP \\
$J$ ~[mag] &  7.646 $\pm$ 0.03  & 2MASS \\

\hline

\noalign{\smallskip}
\multicolumn{3}{c}{\textit{Stellar parameters}} \\
\noalign{\smallskip}
$L_{\rm X}$ ~[$\times 10^{28}$ erg s$^{-1}$] & $1.89\pm0.07$ & Sect.\,\ref{Sect: X observations} \\

$L_{\star}$ ~[$L_{\odot}$]  & 0.3609 $\pm$ 0.0052 & Sect.\,\ref{subsect: Stellar parameters} \\
$T_{\mathrm{eff}}$ ~[K]  & 5037$\pm$14 & Sect.\,\ref{subsect: Stellar parameters} \\
$\log( g\,[$cm\,s$^{-2}] )$ & 4.63 $\pm$ 0.02 & Sect.\,\ref{subsect: Stellar parameters} \\
$\rm [Fe/H]$ & $-$0.01$\pm$0.02 & Sect.\,\ref{subsect: Stellar parameters} \\
  &  $-$0.08$\pm$0.13 & B18 \\
$R_{\star}$ ~[R$_{\odot}$] &  0.789$^{+0.022}_{-0.021}$ & Sect.\,\ref{subsect: Stellar parameters} \\
$M_{\star}$ ~[M$_{\odot}$] &  0.843$^{+0.033}_{-0.056}$ & Sect.\,\ref{subsect: Stellar parameters} \\

$V_{\rm broad}$ ~[km\,s$^{-1}$] &  2.89 $\pm$ 0.03 & Sect.\,\ref{subsect: Stellar parameters} \\
$v\,\sin{i}$ ~[km\,s$^{-1}$] &  $<$2.90 & Sect.\,\ref{subsect: Stellar parameters} \\
$P_{\mathrm{rot}}$ ~[d] & 12.0 $\pm$ 0.4 & Sect.\,\ref{subsect: TESS photometry} \\
$\log( $R$^{'}_{HK} )$ & $-$4.242 $\pm$ 0.016 & M22 \\
Age ~[Myr] & 600--800 & Sect.\,\ref{Sect: Age} \\

\noalign{\smallskip}
\hline
\end{tabular}
}

\tablebib{
HD: \citet{HD_Catalog};
\textit{TESS}: \textit{TESS} Input Catalog v8.2 \citep{TIC_Catalog_Stassun2018};
\textit{Gaia}\,EDR3: \citealp{GAIA_eDR3};
\textit{Gaia}\,DR2: \citealp{GAIA_DR2};
TOI: \citet{Guerrero_2021};
HIP: \citet{HIP_Catalog};
TYC: \citet{TYC_Catalog};
2MASS: \cite{2MASS_Catalog};
B18: \citet{Bochanski2018_metalicity};
M22: \citet{Maldonado_2022}.
}

\tablefoot{
$^{(a)}$ Systemic radial velocity
}

\end{table}

\subsection{CARMENES observations and analysis}

A single transit of the planet candidate \hd23.01 was observed with the CARMENES\footnote{Calar Alto high-Resolution search for M dwarfs with Exoearths with Near-infrared and optical \'Echelle Spectrographs.} (\citealp{Quirrenbach_2014, Quirrenbach_2020}) spectrograph located at the Calar Alto Observatory, Almer\'ia, Spain, on the night of 6 August 2022. CARMENES has two spectral channels: the optical channel (VIS), which covers the wavelength range of 0.52--0.96\,$\mu$m with a resolving power of $\mathcal{R}$\,=\,94\,600, and the near-infrared channel (NIR), which covers 0.96--1.71\,$\mu$m with a resolving power of $\mathcal{R}$\,=\,80\,400. We observed the target with both channels simultaneously and collected a total of 44 spectra of 5\,min exposure time, with 28 of them between the first ($T_1$) and fourth ($T_4$) transit contacts. We obtained a median  S/N of 72 around the H$\alpha$ line and of 86 around the \ion{He}{I} triplet.

Fiber A was used to observe the target star, while fiber B was placed on sky, separated by 88\,arcsec in the east-west direction. The observations were reduced using the CARMENES pipeline \texttt{caracal} \citep{Caballero2016} and both fibers were extracted with the flat-optimized extraction algorithm (\citealp{FOX_extraction}). We also processed the spectra with \texttt{serval}\footnote{\url{https://github.com/mzechmeister/serval}} \citep{SERVAL}, which is the standard CARMENES pipeline to derive the radial velocities (RVs) and several activity indicators: the chromatic radial velocity index (CRX) and the differential line width (dLW), as well as the H$\alpha$, \ion{Na}{I}\,D1 and D2, and \ion{Ca}{II}\,IRT line indices.

We corrected the VIS and NIR spectra from telluric absorptions with \texttt{molecfit} (\citealp{molecfit_1,molecfit_2}). We analyzed the spectroscopic observations via the well-established transmission spectroscopy technique (e.g. \citealp{Wyttenbach_2015, Nuria_2017}). We computed the H$\alpha$ transmission spectra (TS) following the standard procedure. However, because there are OH emission lines from the Earth's atmosphere close to the \ion{He}{I} triplet lines, we applied an extra step before computing the \ion{He}{I} TS.
First, we planned the observations to avoid a complete overlap or superposition of the OH telluric lines and the \ion{He}{I} planetary trace.
We corrected the fiber A spectra from OH telluric emission using fiber B information, which is used to generate an OH emission model for correcting the science spectra. This methodology is based in previous \ion{He}{I} studies with CARMENES (\citealp{Nortmann_WASP-69_He, Salz_2018, AlonsoFloriano_2019, Enric_2020, Nuria_WASP76, Czesla_2021}). In particular, we followed the procedure previously applied in \cite{Orell2022}. Figure\,\ref{Fig: telluric simulation} compares our prediction for the telluric contamination of the \ion{He}{I} triplet lines with the real observations.

\subsection{X-ray observations and planetary irradiation}
\label{Sect: X observations}

We used XMM-Newton archival observations of \hd23 (PI M.\,Zhang) to calculate the X-ray luminosity of the star. The star was observed on 7 July 2021. We reduced the data following standard procedures and used the three EPIC detectors to extract a spectrum for each of them, simultaneously fitting them with a two-temperatures coronal model, using the ISIS package \citep{isis} and the Astrophysics Plasma Emission Database \citep[APED,][]{aped} v3.0.9. A value of interstellar medium (ISM) absorption H column density of 1\,$\times$\,10$^{19}$\,cm$^{-3}$ was adopted, consistent with the fit to the overall spectrum, and the distance to the source. The resulting model has $\log T_{1,2}$\,(K)\,=\,$6.53^{+0.14}_{-0.13}$, $6.87^{+0.07}_{-0.09}$; $\log$\,{\it EM}$_{1,2}$\,(cm$^{-3}$)\,=\,$50.44^{+0.32}_{-0.31}$, $50.64^{+0.18}_{-0.40}$; and abundances $\rm [Fe/H]$\,=\,$-0.24\pm0.17$ and $\rm [Ne/H]$\,=\,0.17$^{+0.28}_{-0.69}$.
An X-ray luminosity $L_{\rm X}$, in the energy range of 0.12--2.48 keV ($\lambda$=5--100\,$\AA$), of $(1.89\pm0.07)\times10^{28}$\,erg\,s$^{-1}$ was calculated. We extrapolated the coronal model towards transition region temperatures, following \citet{san11}, to determine the expected EUV stellar emission. We calculate $L$\,=\,15$^{+15}_{-5} \times 10^{27}$\,erg\,s$^{-1}$ and 7.8$^{+2.9}_{-1.8} \times 10^{27}$\,erg\,s$^{-1}$ for the EUV spectral ranges 100--920\,$\AA$ and 100--504\,$\AA$, respectively. Our X-ray flux is $\sim$30\% lower than the value calculated by \citet{Zhang_young_planets}, likely related to the spectral fitting process (we include the Ne abundance in the fit, and they assumed no ISM absorption, M. Zhang, priv. comm. 2023). \citet{Zhang_young_planets} report also the incident flux (27\,000\,erg\,s$^{-1}$\,cm$^{-2}$) in the range $\lambda \lambda\, 1230-2588$\,\AA\ (defined as MUV in their work), after the modeling of the stellar emission that was then scaled up to match the observed flux in the XMM-Newton Optical Monitor (OM) filters  UVW2 ($\lambda \lambda\, 2120 \pm 500$\,\AA) and UVM2 ($\lambda \lambda\, 2310 \pm 480$\,\AA).
We find an excellent agreement between our modelled SED and the observed count rate in these two filters. However these filters have a non-negligible sensitivity at longer wavelengths. We must remark that the SED of a late-type star, as in our case, yields a $\sim$90\% result for the count rate in the filter actually coming from longer wavelengths than the nominal band-pass (see Appendix\,\ref{App: MUV flux}). Thus, scaling the MUV level based on the photometry of the UV filters on board XMM-Newton must be taken with care for this type of star. In any case, our model (see Sect.\,\ref{sect:model}) indicates an incident flux in the MUV range of 13\,300\,erg\,s$^{-1}$\,cm$^{-2}$, without scaling, based on the OM flux.

\subsection{TESS observations}
\label{Sect: TESS observations}

Listed as TIC 293954617 in the \textit{TESS} Input Catalog (TIC; \citealp{TIC_Catalog_Stassun2018}), \hd23 was observed by \textit{TESS} in 2-min short-cadence integrations in Sectors\,14--16, 41, and 54--56. In particular, the transit studied in this work was simultaneously observed by \textit{TESS} and CARMENES in Sector\,55.


We fit all the \textit{TESS} simple aperture photometry (SAP; \citealp{SAP}), which is publicly available at the Mikulski Archive for Space Telescopes (MAST\footnote{\url{https://mast.stsci.edu/portal/Mashup/Clients/Mast/Portal.html}}), using \texttt{juliet}\footnote{\url{https://juliet.readthedocs.io/en/latest/index.html}} \citep{juliet} to refine the central time of transit ($t_0$) and other planetary parameters, such as the orbital period and the planetary radius.
\texttt{Juliet} is a \texttt{python} library based on other public packages for transit light curve (\texttt{batman}, \citealp{batman}) and GP (\texttt{celerite}, \citealp{celerite}) modelling, which uses nested sampling algorithms (\texttt{dynesty}, \citealp{dynesty}; \texttt{MultiNest}, \citealp{MultiNest, PyMultiNest}) to explore the parameter space. In the fitting procedure, we adopted a quadratic limb-darkening law with the ($q_1$,$q_2$) parameterization introduced by \citet{Kipping2013}, and we considered the uninformative sample ($r_1$,$r_2$) parameterization introduced in \citet{Espinoza2018} to explore the impact parameter of the orbit ($b$) and the planet-to-star radius ratio ($p$\,$=$\,${R}_{\mathrm{p}}/{R}_{\star}$) values. According to the flux contamination exploration in \citet{Zhang_young_planets}, we can safely fix the dilution factor to 1. Furthermore, the \textit{TESS} apertures used to compute each sector light curve only include our target star (see Fig.\,\ref{Fig: TESS_FOV}).
The results from the \textit{TESS} data analysis are presented in Sect.\,\ref{subsect: TESS photometry}.


\subsection{MuSCAT2 observations and data reduction}

We observed a full transit of \hd23.01 on 23 June 2021 with the multi-color imager MuSCAT2 \citep{Narita19}, mounted on the Carlos S\'anchez Telescope (TCS) at Teide Observatory (OT). MuSCAT2 obtained simultaneous photometry in four independent CCDs in the \textit{g}, \textit{r}, \textit{i}, and \textit{$z_{s}$} photometric bands, with an exposure time setting of 5\,s for all CCDs. We performed the data reduction, and the aperture photometry using the MuSCAT2 pipeline described by \cite{Parviainen20}. The multi-color lightcurves (App.\,\ref{App: multicolor_validation}) were computed through a global optimization  that accounts for the transit and baseline variations simultaneously, using a linear combination of covariates.

\section{Stellar characterization}

In this section, we revise the planet host star parameters taking advantage of the high-S/N co-added spectrum from \texttt{serval} and other available photometric data. Table\,\ref{table - System Param} compiles the stellar parameters derived in this work, and from the literature.

\subsection{Stellar parameters}
\label{subsect: Stellar parameters}

\begin{figure}[ht!]
\includegraphics[width=1\linewidth]{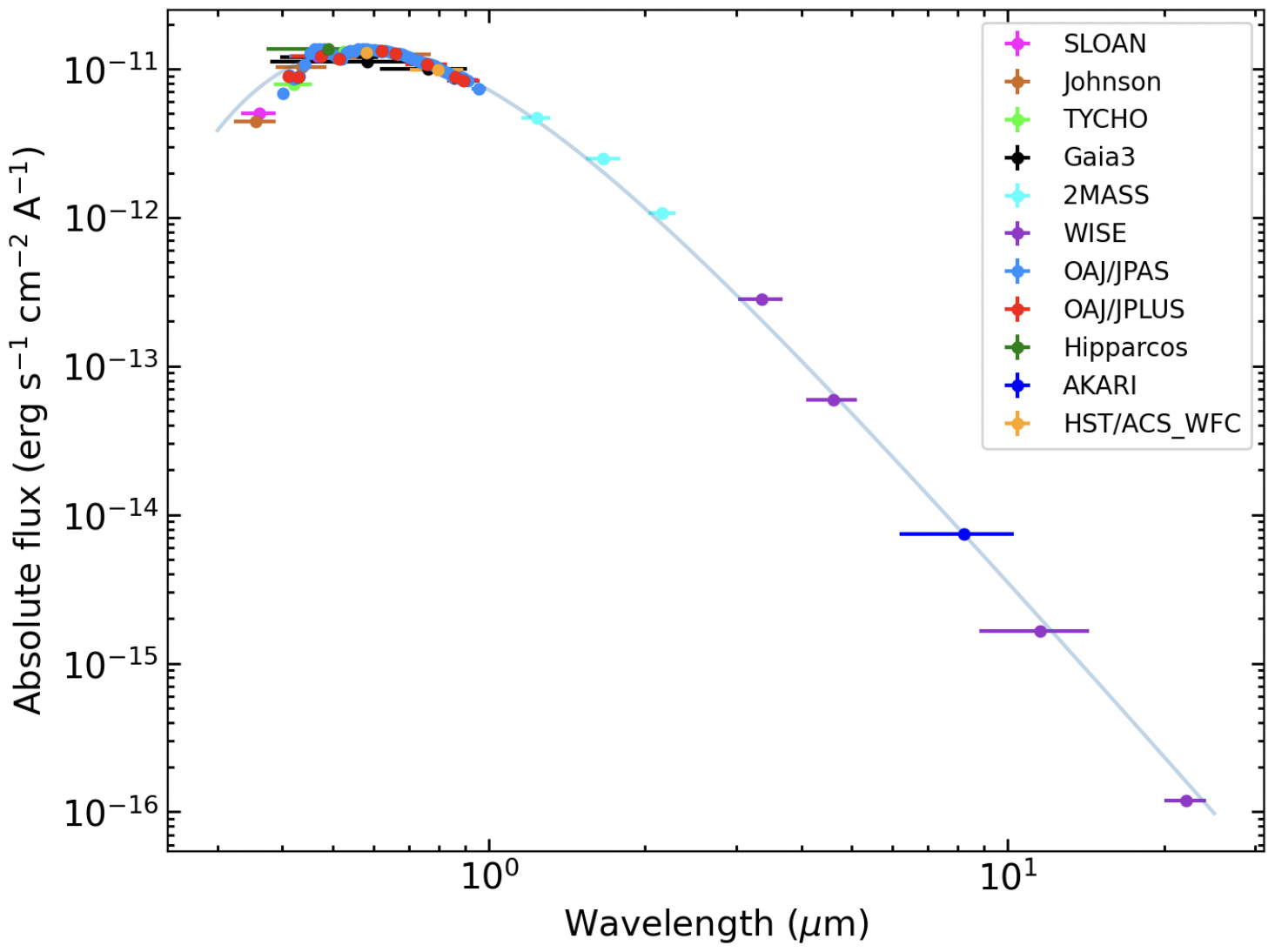}
\caption{Photometric spectral energy distribution of HD\,235088 (circles) from 0.3 through $\sim$25 $\mu$m. A black-body source with a temperature of 5050 K (solid line) is added for the sole purpose of demonstrating that all fluxes are photospheric in origin and there are no infrared flux excesses at long wavelengths. Vertical error bars denote flux uncertainties and horizontal error bars account for the width of the photometric passbands.
\label{Fig:phot_sed}}
\end{figure}

To determine the bolometric luminosity of HD\,235088, we first built the star's photometric spectral energy distribution (SED) using broad- and narrow-band photometry from the literature. The stellar SED is shown in Figure\,\ref{Fig:phot_sed} including the Johnson $UBVR$ photometry \citep{2002yCat.2237....0D}, the $ugriz$ data from the Sloan Digital Sky Survey \citep{2000AJ....120.1579Y}, {\sl Gaia} Early Data Release 3 and Tycho photometry \citep{2022A&A...667A.148G, 2000A&A...355L..27H}, the Two Micron All Sky Survey (2MASS) near-infrared $JHK_s$ photometry \citep{skrutskie2006}, the Wide-field Infrared Survey Explorer ({\sl WISE}) $W1$, $W2$, $W3$, and $W4$ data \citep{wright2010}, the {\sl AKARI} 8-$\mu$m flux \citep{2006PASJ...58..673M}, and the optical multi-photometry of the Observatorio Astrofísico de Javalambre (OAJ), Physics of the Accelerating Universe Astrophysical Survey (JPAS), and Photometric Local Universe Survey (JPLUS) catalogs. All of this photometric information is accessible through the Spanish Virtual Observatory \citep{bayo2008}. In total, there are 90 photometric data points defining the SED of HD\,235088 between 0.30 and $\sim$25 $\mu$m. The OAJ/JPAS data cover very nicely the optical region in the interval 0.40--0.95 $\mu$m with a cadence of one measurement per 0.01 $\mu$m. In Figure\,\ref{Fig:phot_sed}, the effective wavelengths and widths of all passbands were taken from the Virtual Observatory SED Analyzer database \citep{bayo2008}. The {\sl Gaia} trigonometric parallax was employed to convert all observed photometry and fluxes into absolute fluxes. The SED of HD\,235088 clearly indicates its photospheric origin since there are no obvious mid-infrared flux excesses up to 25 $\mu$m.

We integrated the SED displayed in Figure~\ref{Fig:phot_sed} over wavelength to obtain the absolute bolometric flux ($F_{\rm bol}$) using the trapezoidal rule. The $\sl Gaia$ $G$-band flux was not included in the computations because the large passband width of the filter 
encompasses various redder and bluer filters. We then applied $M_{\rm bol} = -2.5~{\rm log}\,F_{\rm bol} - 18.988$ \citep{cushing2005}, where $F_{\rm bol}$ is in units of W\,m$^{-2}$, to derive an absolute bolometric magnitude $M_{\rm bol}$ = 5.846 $\pm$ 0.016 for HD\,235088, from which 
we obtained a bolometric luminosity of 
$L = 0.3609 \pm 0.0052$ L$_\odot$. The quoted error bar accounts for the photometric uncertainties in all observed bands and the trigonometric distance error, although photometry contributes most to the luminosity error. The contribution of fluxes at bluer and redder wavelengths not covered by the photometric observations to the stellar bolometric flux is less than 1\%.

Then, we computed $T_{\rm eff}$, $\log{g}$, [Fe/H], and total line broadening \citep[$V_{\rm broad}$, see][]{brew16} by means of the {\sc SteParSyn} code\footnote{\url{https://github.com/hmtabernero/SteParSyn/}} \citep{tab22}. The {\sc SteParSyn} code is a \texttt{python} implementation of the spectral synthesis method that uses the {\tt emcee} MCMC sampler to infer the stellar atmospheric parameters. To perform the spectral synthesis, we employed a grid of synthetic spectra computed with the Turbospectrum \citep{ple12} radiative transfer code and the MARCS stellar atmospheric models \citep{gus08}. The spectral synthesis employed atomic and molecular data gathered from the Gaia-ESO (GES) line list \citep{hei21}. To determine the stellar parameters, we fit a selection of \ion{Fe}{i,ii} lines that are well suited to analyzing FGKM stars \citep{tab22}. Using {\sc SteParSyn}, we computed the following stellar atmospheric parameters: $T_{\rm eff}$\,=\,5037~$\pm$\,14\,K, $\log{g}$~$=$~4.63~$\pm$~0.02 dex, $\rm [Fe/H]$~$=$\,$-$0.01\,$\pm$\,0.02\,dex, and $V_{\rm broad}$~$=$\,2.89\,$\pm$\,0.03\,km\,s$^{-1}$. Regarding its spectral type, the spectrum of \hd23 is highly similar to that of HD\,166620, which in turn was analyzed by \cite{Marfil_2020} with exactly the same CARMENES configuration as in our observations. In all, HD\,166620 is a well investigated K2\,V standard star with $T_{\rm eff}$\,=\,5039$\pm$\,85\,K and $\log{g}$~$=$~4.66~$\pm$~0.21~dex \citep[see][]{Marfil_2020}, compatible at the $1\sigma$-level to those of \hd23 (e.g. \citealp{1953ApJ...117..313J, 1984ApJ...279..763N, 2004A&A...418..989N}).

The radius of HD\,235088 can be obtained from the Stefan-Boltzmann law that relates bolometric luminosity, effective temperature ($T_{\rm eff}$), and stellar size. From the spectral fitting of the CARMENES data, HD\,235088's $T_{\rm eff}$ is determined to be 5037\,$\pm$\,14\,K. Therefore, we estimate a radius of $R_{\star}$\,=\,$0.789^{+0.022}_{-0.021}$\,R$_\odot$, where the quoted error accounts for the luminosity uncertainty and an increased $\pm$50\,K error in temperature (the latter accounts for the {\sc SteParSyn} error and possible systematics not included in the spectral fitting analysis, e.g. different model atmospheres). This radius determination is independent of any evolutionary model and depends only on the distance (well known with {\sl Gaia}), the bolometric luminosity (well determined), and the model atmospheres used to fit the observed spectra.

The mass of HD\,235088 can be derived following empirical mass--luminosity relationships. \citet{2021A&A...647A..90F} used masses and radii of eclipsing binaries with FGK main-sequence components to obtain a reliable mass--radius relation, from which we inferred a mass of $M_{\star}$\,=\,$0.843 ^{+0.033}_{-0.056}$\,M$_\odot$ for our star (by adopting a solar chemical composition).

\subsection{Stellar rotation and age determination}
\label{Sect: Age}



\cite{Zhang_young_planets} computed a rotational period ($P_{\mathrm{rot}}$) of 5.79\,$\pm$\,0.15\,days from a Lomb-Scargle periodogram of \textit{TESS} SAP data and estimated an age of 165$\pm$30\,Myr using gyrochonologic relations. \cite{Maldonado_2022} derived a similar $P_{\mathrm{rot}}$ using \textit{TESS} data (6.14\,d), but of 12.8--14\,d with STELLA and REM photometric data. However, they estimated a significantly older age  of 600\,Myr for \hd23.

\begin{figure}[ht!]
\includegraphics[width=1\linewidth]{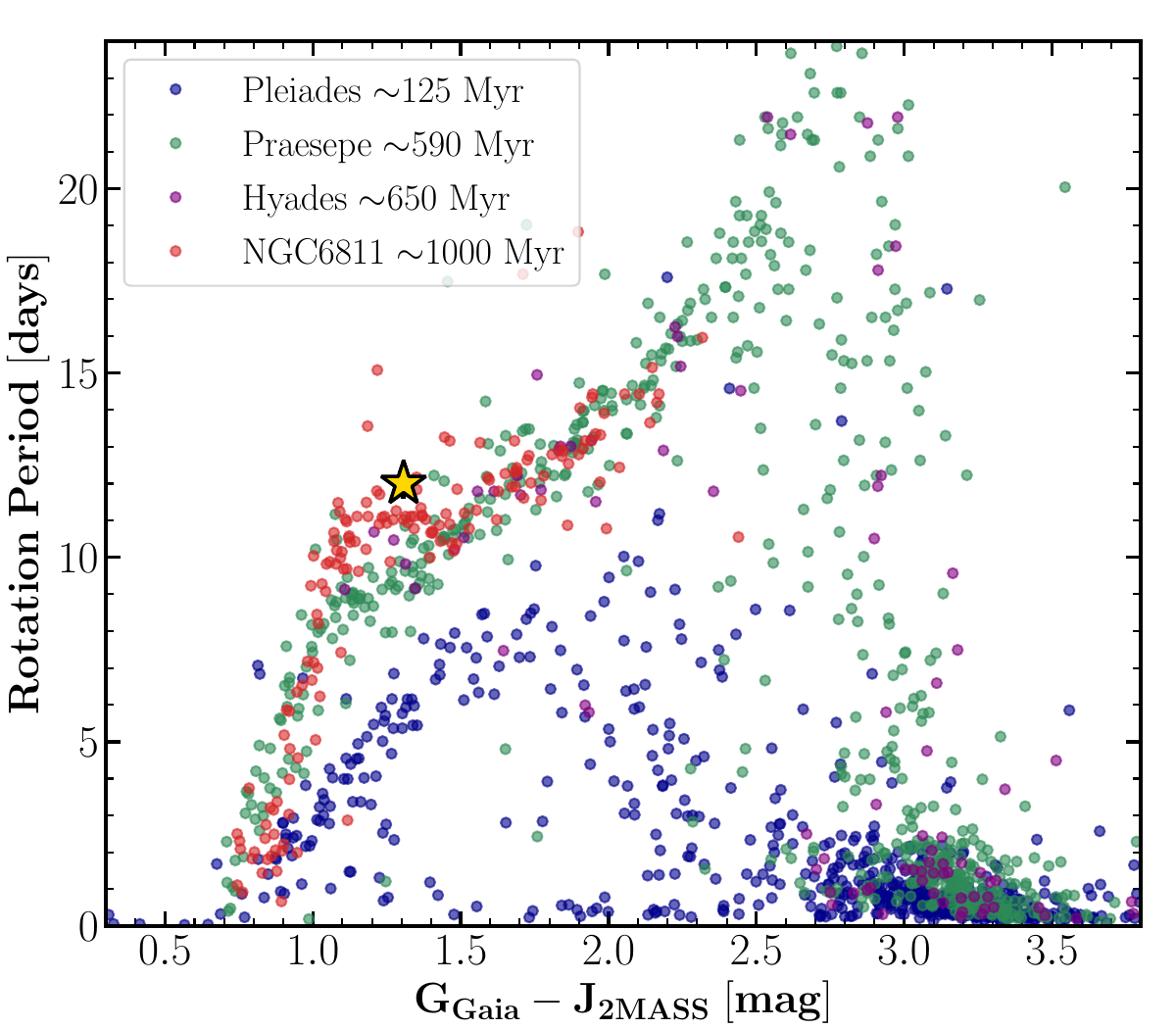}
\caption{Rotation period distribution as a function of color $G$--$J$ for the Pleiades ($\sim$125\,Myr; \citealp{rebull2016}), Praesepe ($\sim$590\,Myr; \citealp{douglas2017}), Hyades ($\sim$650\,Myr; \citealp{douglas2019}) and NGC 6811 ($\sim$1000\,Myr; \citealp{curtis2019}) clusters. The gold star represents \hd23. 
\label{fig:prot_age}}
\end{figure}

\begin{figure}[ht!]
\includegraphics[width=1\linewidth]{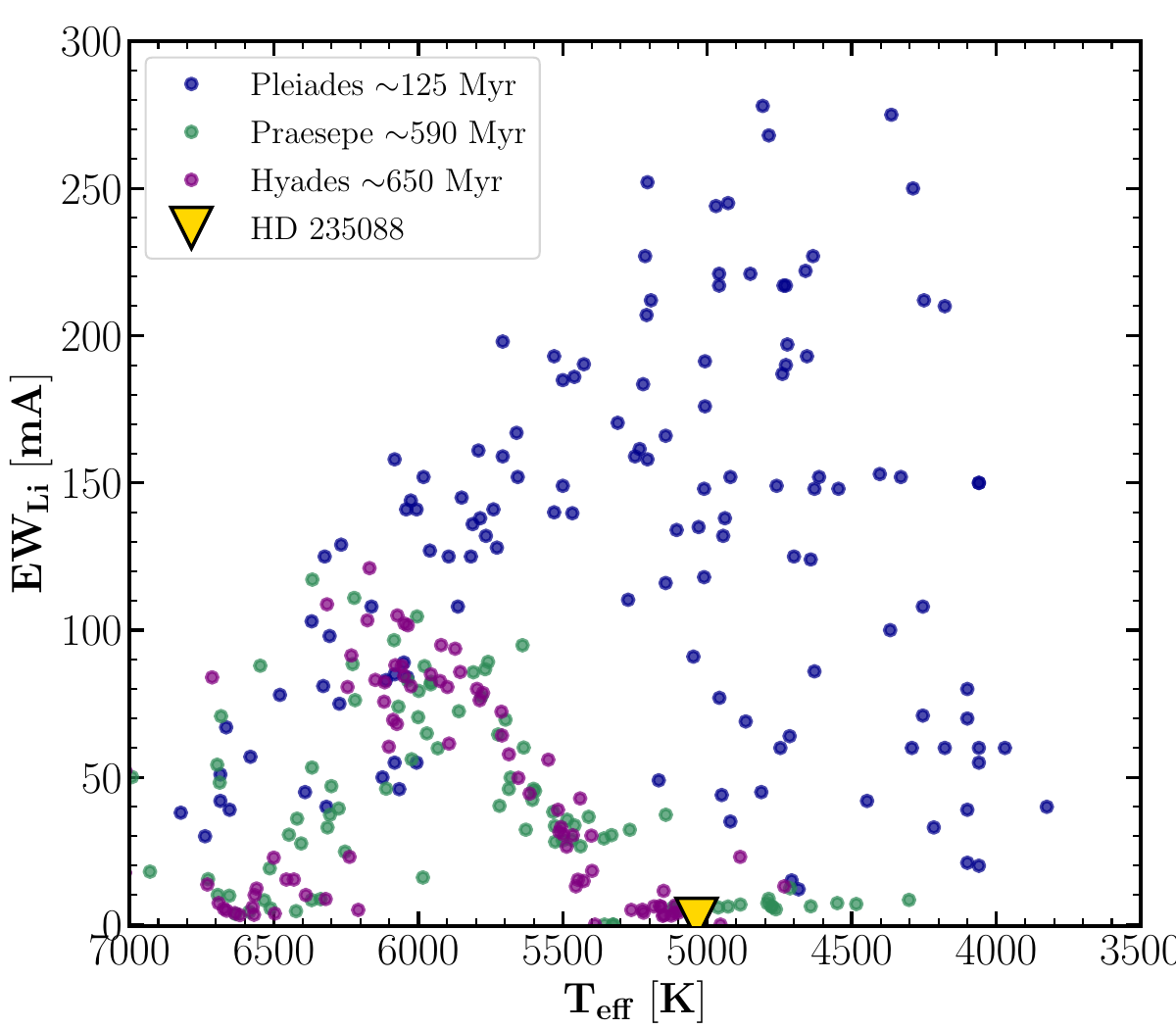}
\caption{Equivalent width distribution of \ion{Li}{i} as a function of the effective temperature for the Pleiades ($\sim$125\,Myr; \citealp{bouvier18}), Praesepe ($\sim$590\,Myr), and Hyades ($\sim$650\,Myr; \citealp{cummings17}). The gold triangle represents \hd23.
\label{fig:ew_li}}
\end{figure}

\begin{figure}[ht!]
\includegraphics[width=1\linewidth]{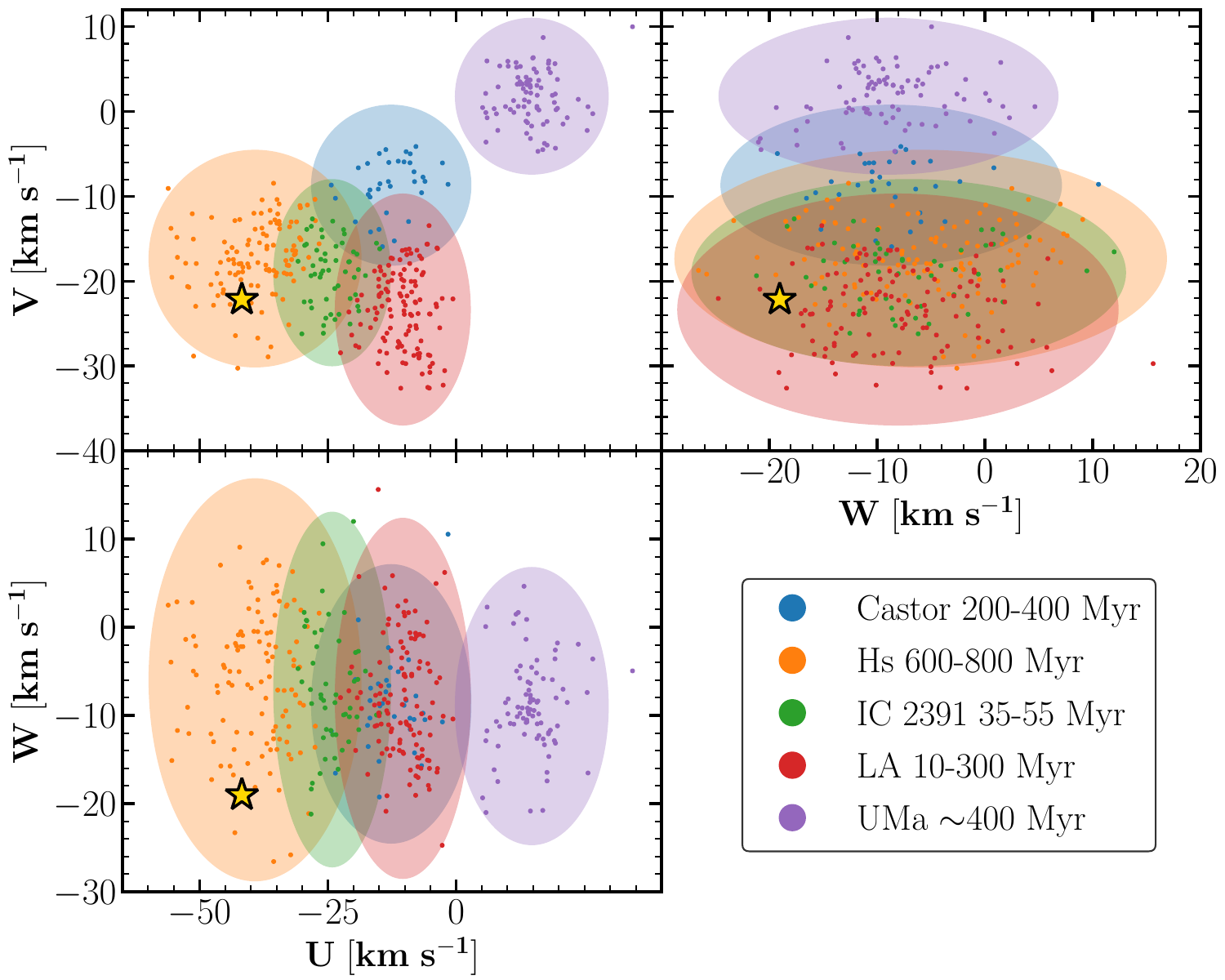}
\caption{$UVW$ velocity diagram for \hd23 (gold star). The members of the Castor moving group, Hyades supercluster (Hs), IC 2391 supercluster, Local Association (LA), and Ursa Major group (UMa) from \cite{montes2001b} are included. The ellipses represent the 3$\sigma$ values of the $UVW$ for each Young Moving Group. \hd23 agrees with the Hyades supercluster.  
\label{fig:UVW}}
\end{figure}

Here, we used seven sectors of \textit{TESS} SAP data to derive a $P_{\mathrm{rot}}$ of 12.0\,$\pm$\,0.4\,days (see Sect.\,\ref{subsect: TESS photometry} for details), which is consistent with the maximum peak at $\sim$11.8\,days from the computed generalized Lomb-Scargle periodogram (\citealp{GLS_paper}, see Fig.\,\ref{fig:GLS_Prot}) and with the X-ray luminosity we measured, according to the relations in \citet{Wright2011}. Quantitatively, we estimated \hd23's age using gyrochronology with the relations presented in \citet[Eqs.\,12, 13, and 14, the parameters from Table\,10]{Mamajek_2008} and \citet[and Eq.\,1]{Schlaufman_2010}, also used in \cite{Zhang_young_planets}. We obtained ages of 630$^{+100}_{-85}$\,Myr and 690$^{+200}_{-80}$\,Myr, respectively. Figure\,\ref{fig:prot_age} shows the distribution of rotation periods as a function of color $G$-$J$ for different young clusters. Qualitatively, the age of \hd23 is consistent with the sequences of Praesepe, Hyades, and NGC 6811, that is, between 590 and 1000\,Myr. Moreover, from the relation between the age for young stars and the coronal X-ray emission \citep[Eq.\,A3]{Mamajek_2008} and using the $L_{\rm X}$ derived in Sect.\,\ref{Sect: X observations}, we obtained 700$^{+1050}_{-425}$\,Myr for \hd23. Another age indicator that is commonly used is the atmospheric absorption of \ion{Li}{I} 6709.61\,\AA. We looked for \ion{Li}{I} in \hd23 in the co-added spectrum generated by \texttt{serval} from the CARMENES spectra, but no clear \ion{Li}{i} feature appeared.
Therefore, we set an upper limit at 3$\sigma$ of 3\,m\AA, indicating that the star is not younger than the Hyades or Praesepe ($\sim$650\,Myr; Fig.\, \ref{fig:ew_li}), namely, the \ion{Li}{i} EW is not compatible with the age proposed by \citet[165\,$\pm$30\,Myr]{Zhang_young_planets}. Lastly, we calculated the $UVW$ galactocentric space velocities \citep{johnson1987} of \hd23 using $Gaia$ astrometry and systemic velocity ($\gamma$) to determine if the object shares kinematics properties with known clusters, moving groups, or associations. 
The $UVW$ velocities of \hd23 are consistent with the young disk and, more particularly, with the Hyades supercluster (Fig.\, \ref{fig:UVW}) which probably indicates its belonging to this supercluster. However, to prove its Hyades supercluster membership a more detailed study must be carried out. From the $UVW$ velocities, we estimated an age of 600-800\,Myr \citep{brant15, lodieu18} which is consistent with the age reported by \cite{Maldonado_2022}, the rotation period, the X-ray emission, and the value we adopted for \hd23.

\section{Validation of the planet candidate}


We used the multi-color transit analysis approach described in \cite{Parviainen19,Parviainen20} to validate \hd23.01 as a planet. The approach uses \texttt{PyTransit} \citep{Parviainen2015} to model the \textit{TESS} photometry with the ground-based multi-color transit photometry to estimate the degree of flux contamination from possible unresolved sources inside the \textit{TESS} photometric aperture. The contamination estimate yields a robust radius ratio estimate for the transiting planet candidate that accounts for any possible third light contamination allowed by the photometry. 
Combining the robust radius ratio with the stellar radius gives an absolute radius estimate of the planet candidate and if this absolute radius can be securely constrained to be smaller than the minimum brown dwarf radius limit ($\sim$0.8\,R$_\mathrm{J}$, \citealt{Burrows2011}), the planet candidate can be considered as a bona-fide transiting exoplanet.
This approach has been used several times in the literature to validate or reject planet candidates orbiting faint stars \citep{Parviainen20,Parviainen21,EsparzaBorges22,Morello22,Murgas22}.

Our analysis, which combines the \textit{TESS} and ground-based MuSCAT2 light curves, gives a robust radius estimate of $R_\mathrm{true}$\,=\,$2.63^{+0.91}_{-0.61}$\,R$_{\oplus}$. This estimate can be considered the most reliable radius estimate for the transiting object when assuming that the \textit{TESS} aperture could contain flux contamination from an unknown (unresolved) source and is consistent with the radius estimate derived from \textit{TESS} photometry ($R_{\rm p}$\,=\,2.045$\pm$0.075\,R$_{\oplus}$). The radius estimate is significantly below the brown dwarf radius limit of 0.8\,R$_\mathrm{J}$, and, thus, we validate \hd23.01 as a sub-Neptune-sized object \hd23\,b.
Furthermore, this analysis rules out a significant flux contamination from an unresolved source of a different spectral type than the host and,  thus, the basic radius obtained with that assumption can be safely adopted as the planet's radius.

\section{Results}
\label{sect: Results}

\subsection{Refinement of the planetary parameters }
\label{subsect: TESS photometry}

\begin{table}[ht!]
\caption[width=\columnwidth]{
\label{table - TOI-1430 juliet priors and posteriors} Planetary parameters for \hd23\,b. Prior and posterior distributions from the \texttt{juliet} fit, and other derived planetary parameters. Prior labels $\mathcal{U}$, $\mathcal{N}$, $\mathcal{F}$, and $\mathcal{J}$ represent uniform, normal, fixed, and Jeffrey's distribution, respectively.
}
\centering
\resizebox{\columnwidth}{!}{%
\begin{tabular}{lcc}

\hline \hline 
\noalign{\smallskip} 

Parameter & Prior & Posterior \vspace{0.05cm}\\
\hline
\noalign{\smallskip}

$P$ [d] & $\mathcal{N}(7.434162,0.01)$ & 7.4341393\,(62)  \vspace{0.05cm} \\ 
$t_0$\,$^{(a)}$ & $\mathcal{N}(2798.46,0.1)$ & 2798.46351$\pm$0.00055  \vspace{0.05cm} \\ 
\textit{e} (deg) & $\mathcal{F }(0)$ & --  \vspace{0.05cm} \\ 
$\omega$ (deg) & $\mathcal{F }( 90)$ & --  \vspace{0.05cm} \\ 
$r_{1}$ & $\mathcal{U }(0,1 )$ & 0.620$\pm$0.030  \vspace{0.05cm} \\ 
$r_{2}$ & $\mathcal{ U}(0,1 )$ & 0.02375$\pm$0.00057  \vspace{0.05cm} \\ 
$\rho_{\star}$ [kg\,m$^{-3}$]  & $\mathcal{ N }(2496.0,350.0)$ & 2585$^{+180}_{-200}$  \vspace{0.05cm} \\
$\mu_{\textit{TESS}}$ (ppm) & $\mathcal{N }(0.0,0.1)$ & 190$^{+580}_{-610}$  \vspace{0.05cm} \\ 
$\sigma_{\textit{TESS}}$ (ppm) & $\mathcal{J }( 10^{-6}, 10^{6})$ & 450$^{+2}_{-2}$  \vspace{0.05cm} \\ 
$q_{1,\textit{TESS}}$ & $\mathcal{U }(0,1 )$ & 0.12$^{+0.12}_{-0.07}$  \vspace{0.05cm} \\ 
$q_{2,\textit{TESS}}$ & $\mathcal{ U}(0,1 )$ & 0.32$^{+0.30}_{-0.20}$  \vspace{0.05cm} \\ 
GP$_\mathrm{B}$ (ppm) & $\mathcal{J }(10^{-6}, 10^{6})$ & 5.0$^{+1.5}_{-1.1}$  \vspace{0.05cm} \\ 
GP$_\mathrm{L}$ [d]  & $\mathcal{J }(10^{-3}, 10^{3})$ & 20$^{+7}_{-5}$  \vspace{0.05cm} \\ 
GP$_\mathrm{C}$ (ppm) & $\mathcal{J }(10^{-6}, 10^{6})$ & 1310$^{+ 57000}_{-1300}$  \vspace{0.05cm} \\ 
GP$_\mathrm{P_{\mathrm{rot}}}$ [d] & $\mathcal{U }(0,20)$ & 12.0$\pm$0.4  \vspace{0.05cm} \\

\noalign{\smallskip} 
\hline 
\noalign{\smallskip} 
\multicolumn{3}{c}{\textit{ Derived planetary parameters }} \\ 
\noalign{\smallskip} 
$p = {R}_{\mathrm{p}}/{R}_{\star}$  & & 0.02375$\pm$0.00057 \vspace{0.05cm} \\  
$b =(a_{\mathrm{p}}/{R}_{\star}) \cos{ i_{\mathrm{p}} }$  & &  0.430$\pm$0.050 \vspace{0.05cm} \\
$a_{\mathrm{p}}/{R}_{\star}$  & &  19.62$^{+0.45}_{-0.50}$ \vspace{0.05cm} \\
$i_{\mathrm{p}}$ (deg)  & & 88.75$\pm$0.17  \vspace{0.05cm} \\  
$T_{14}$\,$^{(c)}$ [h]  &  & 2.700$\pm$0.025 \vspace{0.05cm} \\  
$T_{12}$\,$^{(c)}$ [min]  &  & 4.60$^{+0.30}_{-0.25}$  \vspace{0.05cm} \\  
$R_{\mathrm{p}}$ [${R}_\oplus$]  & & 2.045$\pm$0.075 \vspace{0.05cm} \\  
$a_{\mathrm{p}}$ [AU]  &  & 0.0720$\pm$0.0026  \vspace{0.05cm} \\  
$T_{\mathrm{eq}}$ [K]\,$^{(b)}$  & &  805$^{+13}_{-12}$  \vspace{0.05cm} \\  
${S}$ [$\mathrm{S}_\oplus$] & &  70$\pm$5  \vspace{0.05cm} \\  
$K_{\star}$ ~[m\,s$^{-1}$] & & $\sim$2.5  \\
$K_{\rm p}$\,$^{(d)}$ ~[km\,s$^{-1}$] & & 105 $\pm$ 5 \\

\noalign{\smallskip}
\hline
\end{tabular}
}

\tablefoot{
$^{(a)}$ Central time of transit ($t_0$) units are BJD\,$-$\,2\,457\,000.
$^{(b)}$ Equilibrium temperature was calculated assuming zero Bond albedo.
$^{(c)}$ $T_{14}$ is the total transit duration between the fist ($T_1$) and fourth ($T_4$) contacts, and $T_{12}$ is the duration of the ingress between the first and second contacts.
$^{(d)}$ Calculated from $K_{\rm p} = 2 \pi \, a_{\rm p} \,P^{-1} \sin{i_{\rm p} }$ using the parameters in this table.}
\end{table}

\begin{figure}[ht!]
    \centering
    \includegraphics[width=\hsize]{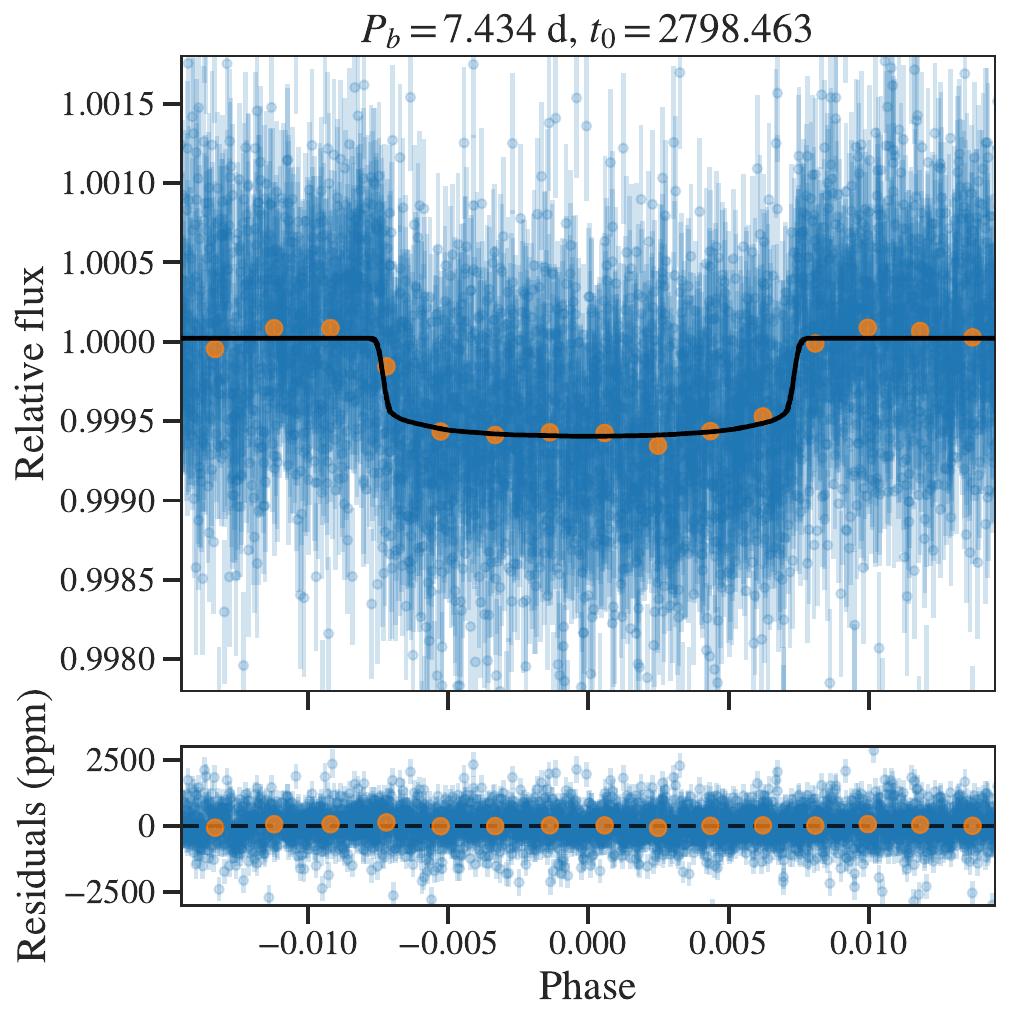}
    \caption{\label{Fig: TOI-1430 phase folded}
    \hd23 \textit{TESS} photometry (blue dots with error bars) phase-folded to the period, $P,$ and central time of transit, $t_0$, (shown above each panel, $t_0$ units are BJD\,$-$\,2\,457\,000) derived from the \texttt{juliet} fit. The black line is the best transit model for \hd23\,b. Orange points show binned photometry for visualization. The GP model was removed from the data.
    }
\end{figure}

We refined the properties of the \hd23 system fitting with \texttt{juliet} the available \textit{TESS} SAP photometric data. We added to the transiting model a \texttt{celerite} GP with quasi-periodic kernel to account for the photometric variability of young star. We adopted the stellar parameters derived in Sect.\,\ref{subsect: Stellar parameters}  and presented in Table\,\ref{table - System Param}. The fitted parameters with their prior and posterior values, and the derived parameters for \hd23\,b are shown in Table\,\ref{table - TOI-1430 juliet priors and posteriors}. The \textit{TESS} data along with the best transiting and GP models are shown in Fig.\,\ref{Fig: TOI-1430 TESS fit}, and \hd23\,b phase folded photometry is shown in Fig.\,\ref{Fig: TOI-1430 phase folded}.

We could improve the ephemeris and properties of \hd23\,b with the new \textit{TESS} data, deriving a radius of $R_{\mathrm{p}}$\,=\,2.045$\pm$0.075\,R$_\oplus$. Moreover, the hyperparameter GP$_{P_{\rm rot}}$ accounts for the periodicities in the photometric variation, which are expected to come from the stellar rotation of the young host star. We obtained GP$_{P_{\rm rot}}$\,=\,12.0\,$\pm$0.4\,days, therefore, we adopt the GP$_{P_{\rm rot}}$ value as \hd23's $P_{\rm rot}$. In Sect.\,\ref{Sect: Age}, we computed with the generalized Lomb-Scargle periodogram method a consistent periodicity ($\sim$11.8\,days, see Fig.\,\ref{fig:GLS_Prot}).

Some values needed to compute the transmission spectra come from the planetary mass measurement, which is still not available for our target at the time of writing. We used the information, and predicted mass ($\sim$7\,$\pm$\,2\,M$_\oplus$) from \cite{Zhang_young_planets} to estimate a semi-amplitude $K_{\star}$ of $\sim$2.5\,m\,s$^{-1}$.


\subsection{Stellar activity analysis}
\label{subsect: stellar activity}

Before computing the TS of the H$\alpha$ and \ion{He}{I} lines, we looked for stellar activity during the observations, which could compromise or challenge the detection of planetary signals, taking advantage of the simultaneous TESS observations to the CARMENES data. However, because \hd23\,b's transit is shallow, and \hd23 is a young star, \textit{TESS} data from a single transit do not have enough quality to clearly see if \hd23\,b passed in front of active regions or spots. There is only a $\sim$20\,minute event close to mid-transit with higher flux than expected in transit, but that level of variation is consistent with the photometric variations detected in other regions of the \textit{TESS} light curve. 

We also checked the time evolution of the \texttt{serval} activity indicators during the observations and constructed the light curve of several stellar lines (all the wavelengths are given in vacuum), namely: \ion{He}{i}\,D3 (5877.2\,$\AA$), \ion{H}{}\,Paschen\,$\beta$ (Pa-$\beta$, 12821.6\,\AA), \ion{H}{}\,Paschen\,$\gamma$ (Pa-$\gamma$, 10941.1\,\AA), and \ion{H}{}\,Paschen\,$\delta$ (Pa-$\delta$, 10052.1\,\AA). The time evolution of these lines are shown in Fig.\,\ref{Fig: Activity}.
They are mainly flat except for the \texttt{serval} H$\alpha$ index, which exhibits a strange behaviour just at the beginning of the transit. The H$\alpha$ index increases smoothly, coincident with the ingress time, and then decreases at about after mid-transit.
However, the \ion{He}{I}\,D3 light curve is stable during the observations without evidence of fluctuations, suggesting that \ion{He}{I} lines were not extremely affected by the H$\alpha$ variability. This is consistent with the \ion{He}{I} line being significantly less variable than the H$\alpha$ line in the overall M dwarf sample observed by CARMENES (\citealp{Fuhrmeister_2020A&A...640A..52F}).

Because we used fiber B to monitor the sky, there were no simultaneous Fabry-P\'erot calibrations during the observations. Thus, the expected amplitude of $\sim$1\,m\,s$^{-1}$ for \hd23\,b's Rossiter-McLaughlin effect is below the uncertainties of the measured RVs.

\subsection{H$\alpha$ and \ion{He}{I} transmission spectra} \label{tra_spec}

\begin{figure*}
    \centering
    \includegraphics[width=\hsize]{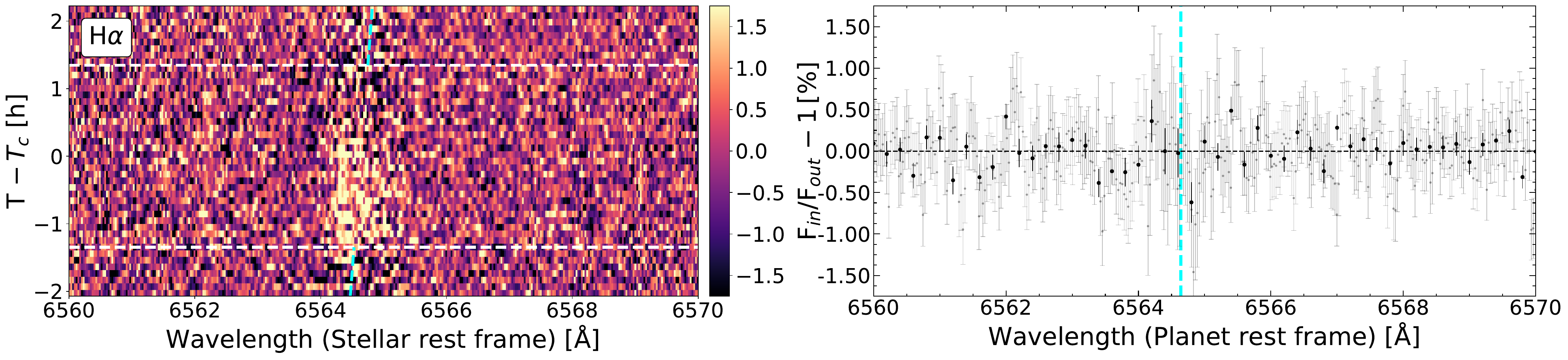}
    \includegraphics[width=\hsize]{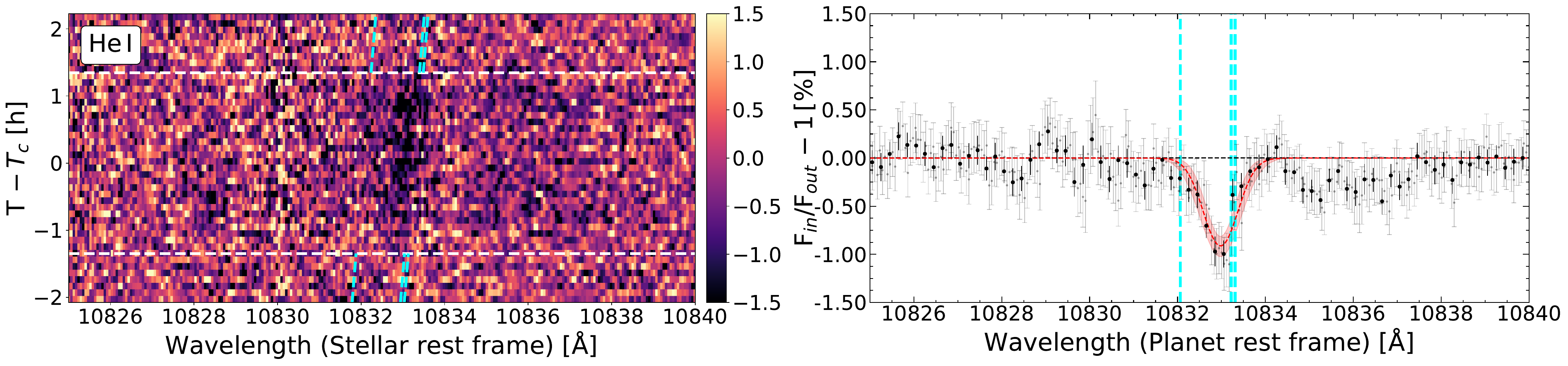}
    \caption{\label{Fig: TS PLOTS} Residual maps and transmission spectra around H$\alpha$ (\textit{top panels}) and \ion{He}{I} NIR (\textit{bottom panels}) lines. \textit{Left panels}: Residual maps in the stellar rest frame. The time since mid-transit time ($T_c$) is shown on the vertical axis, wavelength is on the horizontal axis, and relative absorption is color-coded. Dashed white horizontal lines indicate the first and fourth contacts. Cyan lines show the theoretical trace of the planetary signals. \textit{Right panels}: Transmission spectra obtained combining all the spectra, uncontaminated from stellar activity, between the first and fourth contacts. We show the original data in light gray and the data binned by 0.2\,\AA\ in black. The best Gaussian fit model is shown in red along with its $1\sigma$ uncertainties (shaded red region). Dotted cyan vertical lines indicate the H$\alpha$ (\textit{top}) and the \ion{He}{I} triplet (\textit{bottom}) lines positions. All wavelengths in this figure are given in vacuum.
    }
\end{figure*}

\begin{figure*}
    \centering
   \begin{subfigure}{0.49\textwidth}
         \centering
         \includegraphics[width=\textwidth]{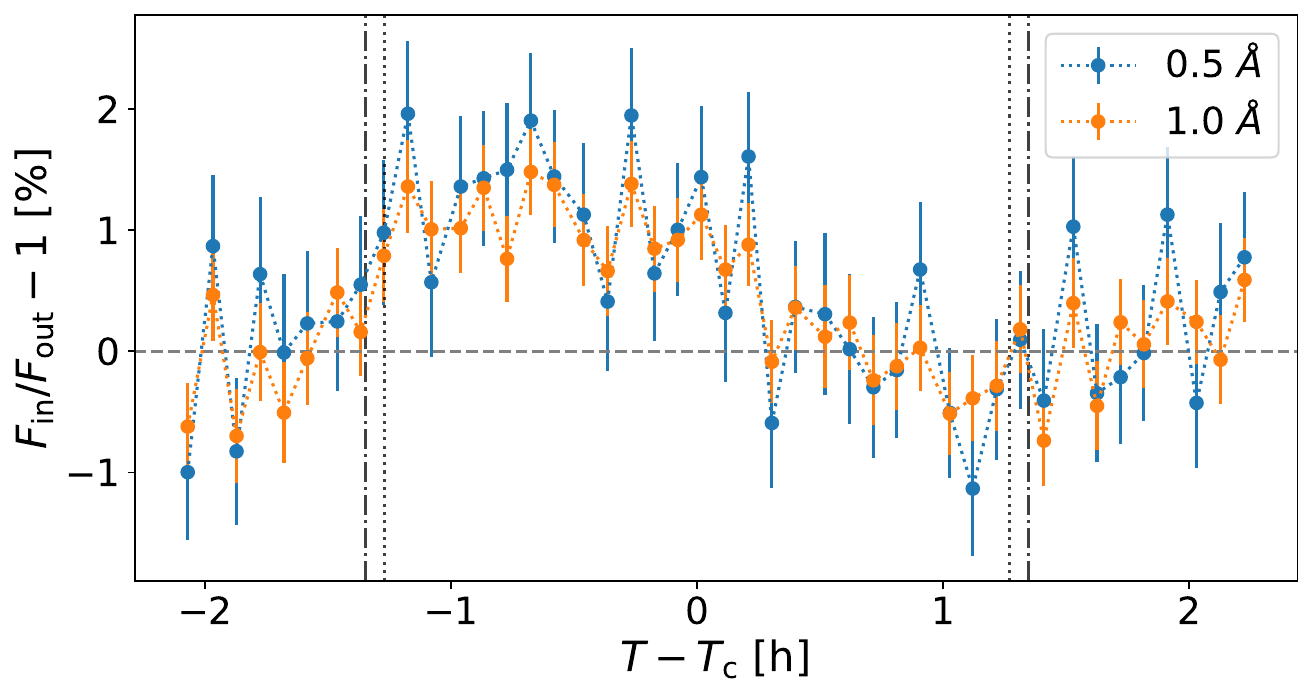}
     \end{subfigure}
     \hfill
     \begin{subfigure}{0.49\textwidth}
         \centering
         \includegraphics[width=\textwidth]{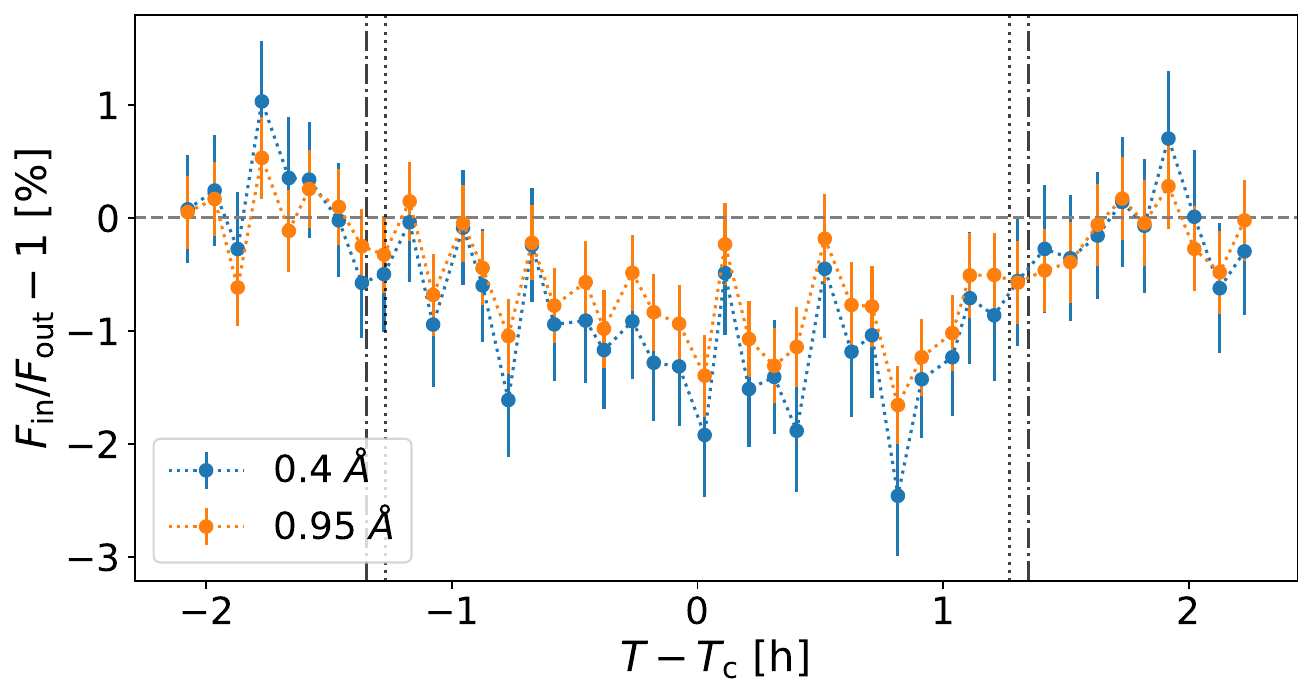}
     \end{subfigure}
    \caption{Transit light curves of H$\alpha$ (\textit{left}) and \ion{He}{I} triplet (\textit{right}). The \ion{He}{I} light curve has been constructed integrating the counts of the residual map in the planet rest frame around $\lambda_0$ using $\sigma$ (blue) and FWHM (orange) wavelength band passes from Table\,\ref{table - TOI-1430 priors and posteriors}. For the H$\alpha$ light curve, we considered two generic band passes. The vertical lines represent the different contacts during the transit. Note the different y-axis scales between plots.}
    \label{Fig: TLC}
\end{figure*}

Figure\,\ref{Fig: TS PLOTS} (top) shows the residual map and TS around the H$\alpha$ line. The residual map displays absorption coincident in time with the H$\alpha$ line index feature. To avoid the affected spectra by H$\alpha$ variability, we computed the H$\alpha$ TS without these spectra. We only included the spectra from the second half of the transit in the calculation of the TS. The H$\alpha$ TS (Fig.\,\ref{Fig: TS PLOTS} top right) is mainly flat without a clear planetary absorption signal. We could only set an upper limit to H$\alpha$ excess absorption of $\sim$0.9\,\%, which is computed as three times the root mean square of the TS near the line of interest.

The \ion{He}{I} triplet residual map in Fig.\,\ref{Fig: TS PLOTS} (bottom) shows an absorption region only during the transit, and consistent with the expected location of the planetary trace. From Sect.\,\ref{subsect: stellar activity}, we conclude that there is no evidence of strong contamination of the \ion{He}{I} lines. Thus, the TS shown in the bottom right panel in Fig.\,\ref{Fig: TS PLOTS} was computed from all the spectra taken fully in-transit, between the first and fourth contacts. The TS shows a clear absorption feature from the two strongest lines of the \ion{He}{I} NIR triplet.

We fit the \ion{He}{I} signal with a Gaussian profile, sampling from the parameter posterior distributions using the MultiNest algorithm (\citealp{MultiNest}) via its {\tt python} implementation \texttt{PyMultinest} \citep{PyMultiNest}. We used uniform priors for the excess absorption between $\pm$3\,\%, and the central position ($\lambda_0$) from 10830\,\AA\ to 10835\,\AA. Table\,\ref{table - TOI-1430 priors and posteriors} summarizes the priors and posteriors from the nested sampling fit, and other derived properties as well. Figure\,\ref{Fig: TOI-1430 nested Helium} displays the posterior corner plot of the distributions.
We obtained an \ion{He}{I} triplet absorption signal of $-$0.91$\pm$0.11\,\%, with an equivalent width of 9.5$\pm$1.1\,m\AA\  and significantly blue-shifted ($-$6.6$\pm$1.3\,km\,s$^{-1}$).

Our absorption is deeper, but consistent at the 2$\sigma$-level with the previous detection with Keck/NIRSPEC reported by \cite{Zhang_young_planets} (depth $-$0.64$\pm$0.06\,\%, equivalent width 6.6$\pm$0.5\,m\AA, and blue shift $-$4.0$\pm$1.4\,km\,s$^{-1}$). If we interpret the mild tension between the two values of the absorption depth as variability, it could be shown to originate from the stellar activity. The stellar activity revealed in our analysis of the H$\alpha$ line could increase the XUV stellar flux, increasing the population of the \het\ level.
\cite{Zhang_young_planets} did not present a study of the stellar activity in other than the \het\ lines during their observations, preventing us from examining whether the different \het\ signals are caused by the stellar variability between the two transits. 
Keck/NIRSPEC \textit{Y} band covers from 0.946--1.130\,$\mu$m, in which there are two lines of the \ion{H}{}\,Paschen series: Pa-$\gamma$ (10941.1\,\AA), and Pa-$\delta$ (10052.1\,\AA). The light curves of the stellar Pa-$\gamma$, Pa-$\delta$, and Pa-$\beta$ as well, are shown in Fig.\,\ref{Fig: Activity}. As we noted in Sect.\,\ref{subsect: stellar activity}, those \ion{H}{}\,Paschen lines do not show evidence of stellar activity or variability, as it is detected in the \texttt{serval} H$\alpha$ index and the H$\alpha$ transit light curve. Thus, it seems the lines in the NIR may not be the best option to check for stellar activity or variability.


The transit light curves (TLC) of individual lines are useful for exploring the time evolution of the emission and absorption features reported in the residual maps and TS. The H$\alpha$ and \ion{He}{I} triplet TLCs are displayed in Figure\,\ref{Fig: TLC}.

We computed the TLC for the planetary H$\alpha$ and \ion{He}{I} triplet lines following the methodology applied in \citet{Orell2022}, based on previous transmission spectrum analyses (i.e. \citealp{Nortmann_WASP-69_He, Nuria_2017, Czesla_2021}). We considered two different band-passes to integrate the counts in the planet rest frame. For the H$\alpha$ TLC, we took the generic values of 0.5\,$\AA$, and 1\,$\AA$ centered at the nominal H$\alpha$ wavelength. For the \ion{He}{I} triplet TLC, the band-passes are equal to the fitted $\sigma$ width (0.4\,$\AA$), and the FWHM width (0.95\,$\AA$), both centered at the fitted $\lambda_0$ to account for the detected blue-shift.

The H$\alpha$ TLC (Fig.\,\ref{Fig: TLC} left) is clearly affected by the stellar variability, and shows a very similar time evolution to the \texttt{serval} H$\alpha$ line index. The data points from the second half of the transit are consistent with a null absorption within 1$\sigma$.

The \ion{He}{I} TLC (Fig.\,\ref{Fig: TLC} right) displays the time evolution of the planetary transit. From the \ion{He}{I} TLC with a band pass of 0.4\,$\AA$, we retrieved an excess transit depth of $-$1.00$\pm$0.11\,\% compared to the continuum, which is consistent with the TS absorption. When using a broader band pass of 0.95\,$\AA$, we get an excess transit depth of $-$0.74$\pm$0.07\,\%. We do not detect the pre-transit absorption reported by \citet{Zhang_young_planets}. The two first points after transit are slightly negative, but consistent with null absorption, but below 0\,\%. Thus, we recomputed the \ion{He}{i} residual map, TS, and TLC again with those points as part of the transit. However, we obtained very similar results, suggesting that there is no clear evidence for an extended \ion{He}{I} tail (\citealp{Nortmann_WASP-69_He, WASP-107_Kirk_Tail, WASP-107_He_tail}).



\begin{figure}[ht!]
    \centering
    \includegraphics[width=\hsize]{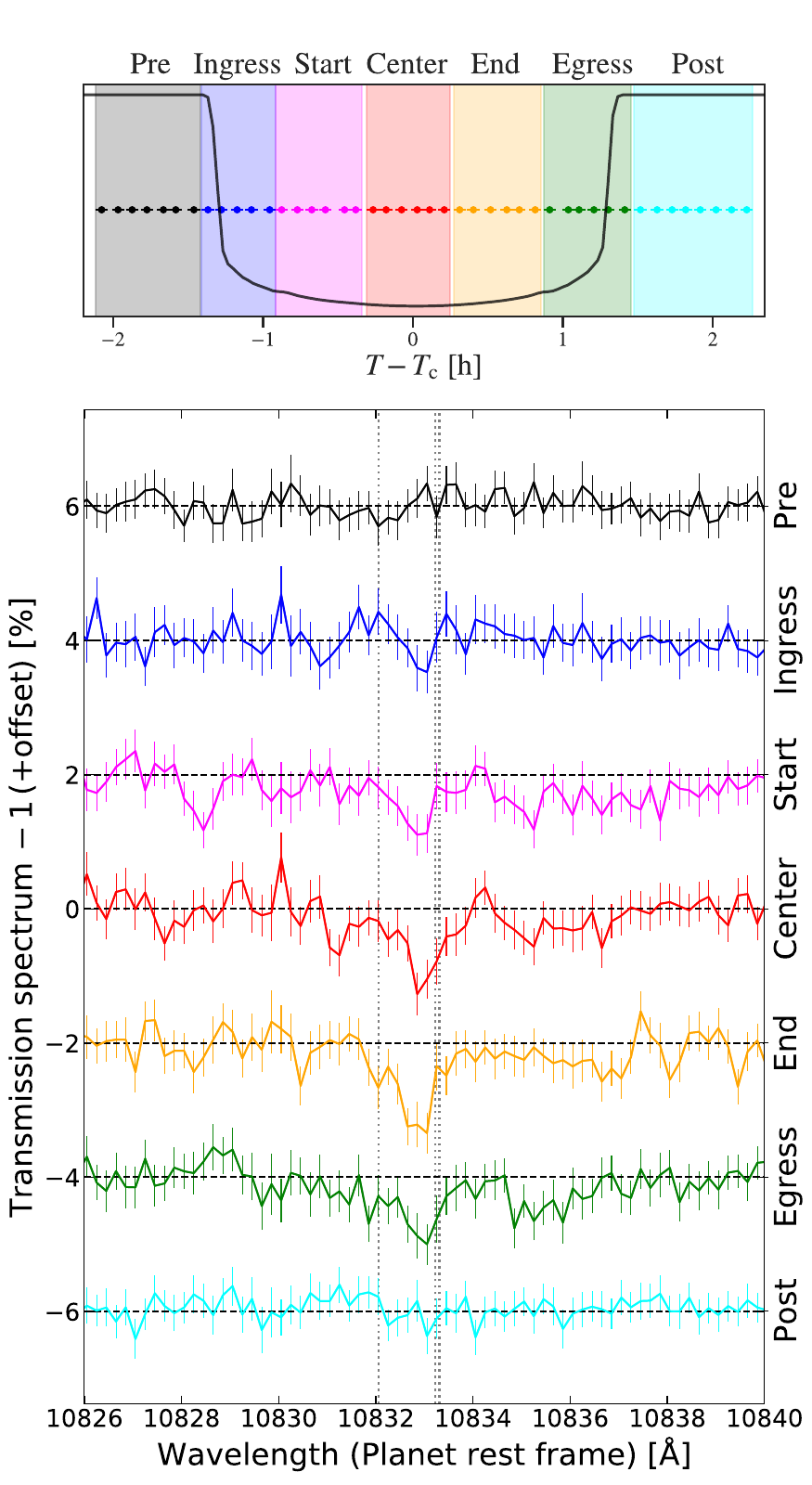}
    \caption{\label{Fig: CARMENES sectors}
    Temporal evolution of the \ion{He}{I} signal.
    \textit{Top panel}: Color scheme for the CARMENES spectra: pre-transit (black), ingress (blue), start (magenta), center (red), end (orange), egress (green), and post-transit (cyan). Horizontal bars represent the 5\,minute duration of the exposures. The transit model for \hd23\,b from \texttt{juliet} (black solid line) is superimposed.
    \textit{Bottom panel}: Transmission spectra (TS) around the \ion{He}{I} triplet line for the transit phases from top to bottom (consecutively offset by 2\,\%). We show each TS binned by 0.2\,$\AA$ with its uncertainties. Dashed black horizontal lines indicate the 0\,\% level in each case, and dotted grey vertical line indicate the \ion{He}{I} triplet line position. The wavelengths in this figure are given in vacuum.}
\end{figure}

\begin{table}[ht!]
\caption[width=\textwidth]{
\label{table - TS and TLC sectors}
Absorption depths retrieved for each out-of- and in-transit phase from transmission spectrum (TS; see Fig.\,\ref{Fig: CARMENES sectors}) and transit light curve (TLC) methods.
}
\centering

\begin{tabular}{lcc}

\hline \hline 
\noalign{\smallskip} 

Phase & TS\,[\%] & TLC\,[\%]\,$^{(a)}$ \vspace{0.05cm}\\
\hline
\noalign{\smallskip}

Pre-transit & $+$0.05$^{+0.30}_{-0.25}$ & $+$0.10$\pm$0.20  \vspace{0.05cm} \\ 
Ingress & 0.00$^{+0.30}_{-0.40}$ & $-$0.08$\pm$0.26  \vspace{0.05cm} \\ 
Start & $-$0.70$^{+0.17}_{-0.23}$ & $-$0.85$\pm$0.17  \vspace{0.05cm} \\ 
Center & $-$0.92$\pm$0.25 & $-$1.10$\pm$0.20  \vspace{0.05cm} \\ 
End & $-$1.21$^{+0.20}_{-0.23}$ & $-$1.33$\pm$0.21  \vspace{0.05cm} \\
Egress & $-$0.76$^{+0.16}_{-0.20}$ & $-$0.76$\pm$0.17  \vspace{0.05cm} \\
Post-transit & $-$0.03$^{+0.23}_{-0.21}$ & $-$0.19$\pm$0.18  \vspace{0.05cm} \\ 

\noalign{\smallskip}
\hline
\end{tabular}

\tablefoot{$^{(a)}$ We computed the TLC with the band-pass equal to the fitted $\sigma$ width, and centered at the $\lambda_0$ from the TS fit for each phase. For pre- and post-transit phases we used a band-pass equal to 0.5\,$\AA$.}
\end{table}

The \ion{He}{I} TLC is asymmetric. Qualitatively, the \ion{He}{I} absorption increases from ingress until mid-transit, and remains constant until the rapid decrease at egress. In a band-pass of 0.4\,$\AA$, the first half and second half of the transit have a transit depth of $-$0.88$\pm$0.14\,\%, and $-$1.15$\pm$0.15\,\%, respectively. We divided the CARMENES spectra in seven different phases to study the temporal evolution of the \ion{He}{I} triplet signal. We defined the phases as \textit{i}) pre-transit, \textit{ii}) around the ingress, \textit{iii}) between ingress and the center of the transit (start), \textit{iv}) center of the transit, \textit{v}) between center of the transit and egress (end), \textit{vi}) around the egress, and \textit{vii}) post-transit. Each in-transit phase has between about five and six spectra, so the results have comparable uncertainties. Figure\,\ref{Fig: CARMENES sectors} (top panel) displays an infographic about the spectra used, and the coverage of each defined phase. We computed the TS, and the TLC for each phase. Figure\,\ref{Fig: CARMENES sectors} displays the TS for each phase, and the absorption values from TS and TLC are presented in Table\,\ref{table - TS and TLC sectors}.
We checked that the TS and TLC \ion{He}{I} signals during the pre- and post-transit phases are consistent with null absorption from the planet. Because the ingress and egress duration of \hd23\,b  is of the order of one exposure (5\,min), it is difficult to inspect the differences between the terminators. At best, we can use our defined ingress and egress phases as a proxy, but these include also spectra that are taken close in time, but not strictly during the ingress and egress, respectively.

We can clearly see differences between the ingress and egress phases. The ingress absorption is consistent with 0\,\%, whereas we computed a significant absorption of $-$0.76\,\% from the egress TS and TLC. Those differences could be produced by variations between the terminators, but could also have their origin in the stellar variability detected in the H$\alpha$ line only during the first half of the transit.
Although we did not, similarly to the work of \cite{Zhang_young_planets},  find any evidence of a cometary-like tail in \hd23\,b, another plausible explanation could be the material accumulation in the egress terminator due to an incipient or failed formation of an \ion{He}{I} tail.
From Fig.\,\ref{Fig: CARMENES sectors}, the \ion{He}{I} signal appears to be consistently blue-shifted during the transit, and it reaches deepest during the center and end phases. The egress phase has an absorption comparable to the start phase one.

\section{Modeling the \ion{He}{I} absorption}\label{sect:model}

\begin{figure}
\centering
\includegraphics[angle=90,width=1.0\columnwidth]{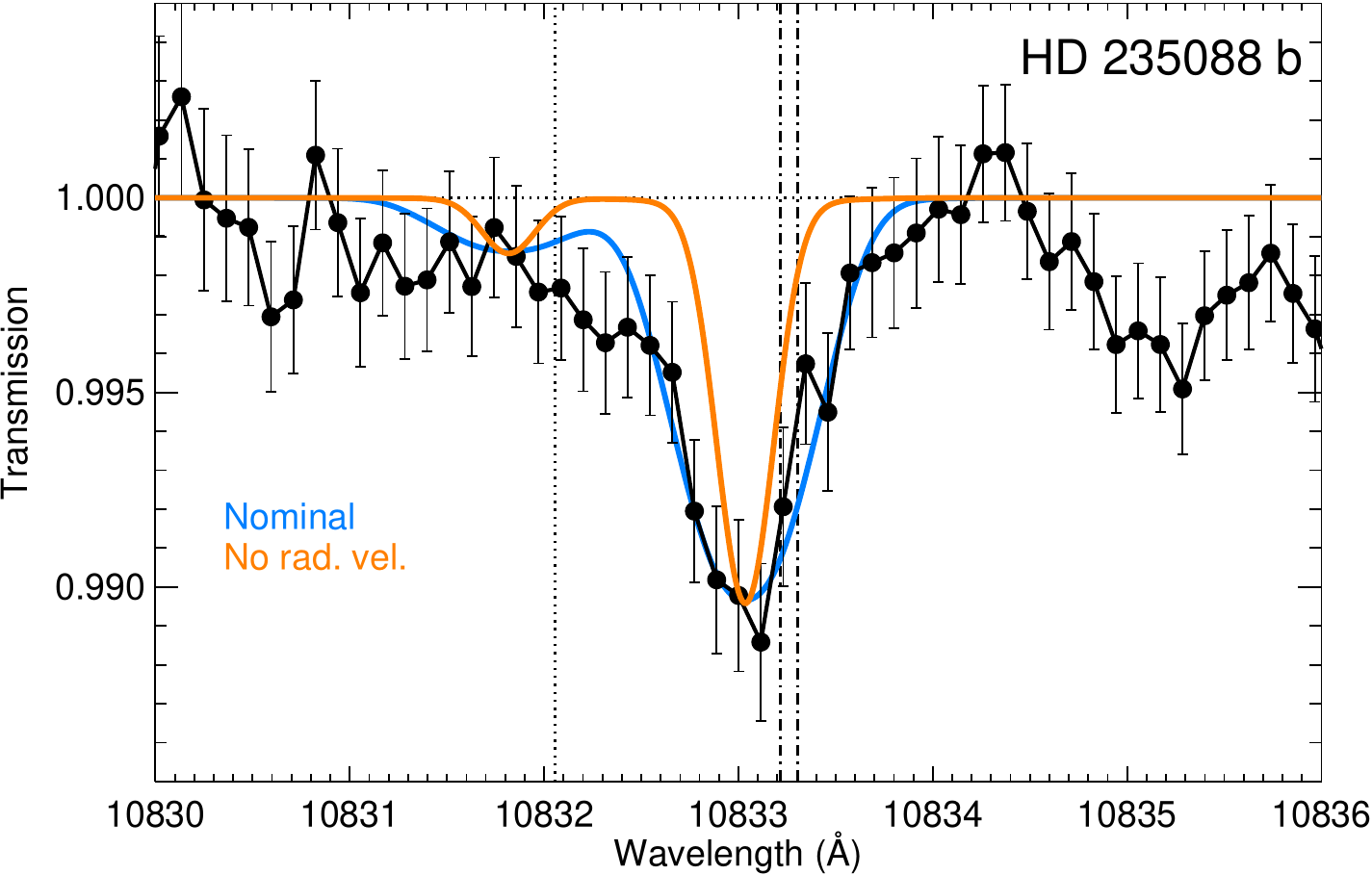}\hspace*{0.35cm}
\caption{Transmission spectrum of the \ion{He}{I} triplet during transit. Measured absorption (solid dots) and estimated errors are shown in black (light gray data in the bottom/right panel of Fig.\,\ref{Fig: TS PLOTS}, but binned to half the number of the points and the error reduced by $\sqrt{2}$). The orange curve shows the absorption profile when only the temperature and turbulence broadenings are included. The blue curve is the best-fit model obtained for an effective temperature of 3000\,K, a mass-loss rate ($\dot{M}$) of 2.4$\times\,10^{10}$\,\gs\, and an H/He ratio of 98/2. The calculation includes a net global wind of $-$6.6\,km\,s$^{-1}$, as derived in Sect.\,\ref{tra_spec}. The positions of the three He~{\sc i} lines are marked by dotted and dash-dotted vertical lines. The wavelengths in this figure are given in vacuum.
}  
\label{model_absorption} 
\end{figure}

\begin{figure}
\includegraphics[angle=90, width=1.\columnwidth]{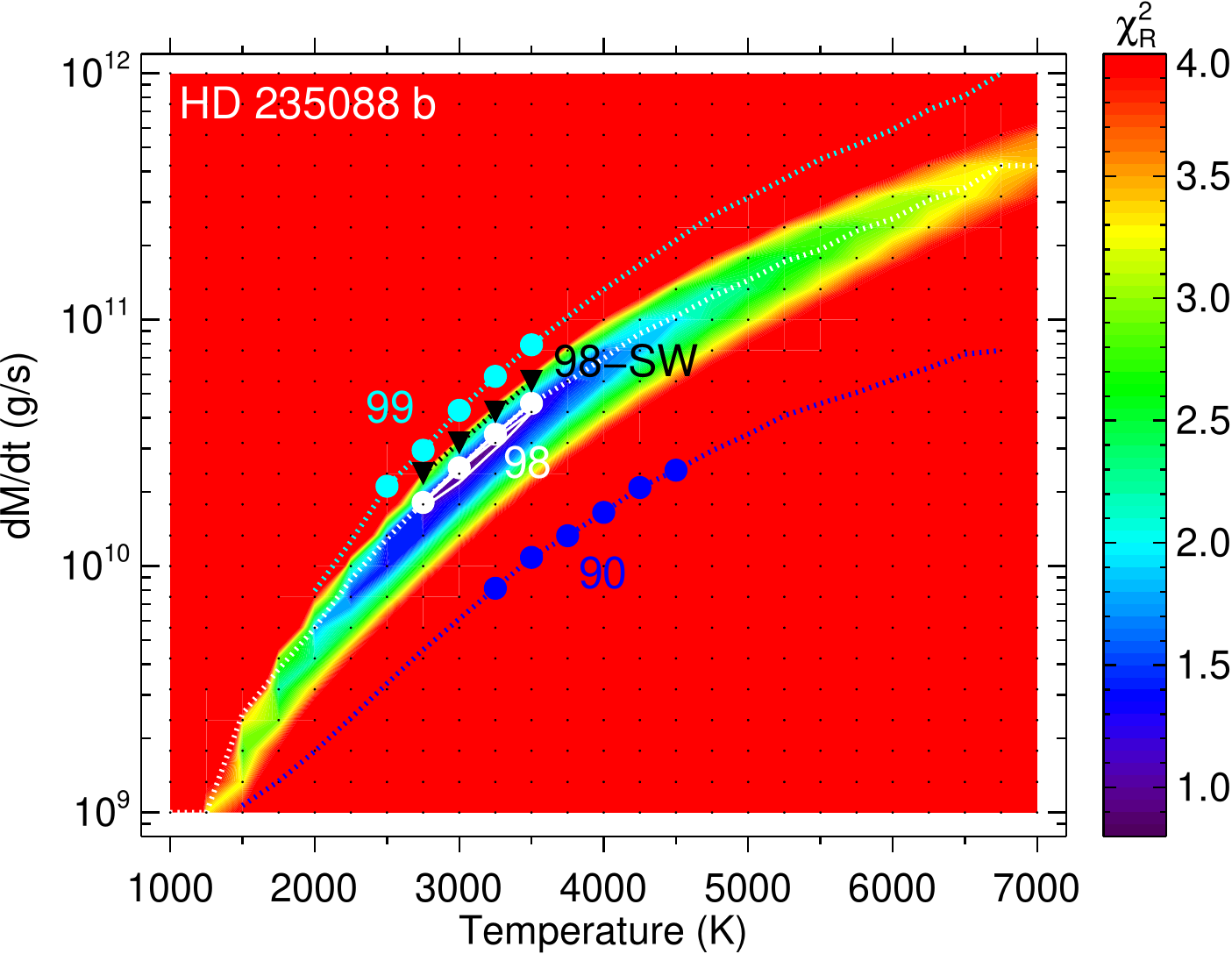} 
\caption{Contour maps of the reduced $\chi^2$ of the \het\ absorption obtained in the modeling for an H/He of 98/2. Dotted curves represent the best fits, with filled symbols denoting the constrained ranges for \mlr\ and temperature for a confidence level of 95\%. 
We over-plotted the curves and symbols when including the estimated effects of the stellar winds (H/He=98/2, downward triangles) and for H/He ratios of 99/1 and 90/10. The labels correspond to the hydrogen percentage, e.g. `98' for an H/He of 98/2. The small black dots represent the $T$-\mlr\ grid of the simulations.}
\label{chi2}
\end{figure}



We analyzed the \ion{He}{I} absorption spectrum following the method used in previous studies for \hdj20, \hdu18, and \gj34 \citep{Lampon2020,Lampon2021a,Lampon2021b}, and (more recently) for HAT-P-32\,b, WASP-69\,b, GJ\,1214\,b, and WASP-76\,b  \citep{Lampon2022}.
Briefly, we used a one-dimensional hydrodynamic and spherically symmetric model, 
together with a non-local thermodynamic equilibrium model to calculate the \het\ density distribution in the upper atmosphere of the planet \citep{Lampon2020}. The \het\ absorption was subsequently computed by using a radiative transfer code for the primary transit geometry \citep{Lampon2020}, which includes Doppler line shapes broadened by the atmospheric temperature, by turbulent velocities and by the velocity of the outflowing gas along the line of sight. We computed the phase averaged synthetic absorption (i.e. the average from $T_1$ to $T_4$ contacts) and included the effects of the impact parameter. Some improvements and updates of the model are described in \citet{Lampon2022}. 
Our estimate of the planetary mass corresponds to a planet-to-star mass ratio of 2.5\,$\times$\,10$^{-5}$. Using the formula given by \citet{Eggleton1983}, we estimate a Roche lobe radius of 2.2\,$R_{\mathrm{J}}$ for the planet, which yields a transit depth of about 8.0\,\% for the Roche volume. The observed 0.90\,\% signal can, therefore, be plausibly caused by material within the planetary Roche lobe.
In our detailed analysis (see below), we found that for an intermediate temperature of the best fits (see Fig.\,\ref{chi2}), the contribution of the layers inside the Roche lobe (under the assumption of no stellar winds, SWs) to the total absorption is about 80\%. This contribution is larger for lower temperatures and smaller for higher temperatures. Also, it is larger if stellar winds are considered.

The model inputs specific to this planet are described below. 
The stellar and planet parameters of the system are listed in Table\,\ref{table - System Param}. 
A key input is the stellar flux from XUV to near-UV, 
covering the range from 1\,\AA\ up to 2600\,\AA. The 1--1600\,\AA\ spectral energy distribution was modeled as described in Sect.\,\ref{Sect: X observations}, and the coverage up to 2600\,\AA\ was comleted by using the photospheric models by \citet{cas03}.
The H/He ratio is another key parameter affecting the mass-loss rate, \mlr,\ and temperature ranges of the upper planetary atmosphere. In previous studies, this ratio has been constrained by using atomic hydrogen absorption measurements (\lya\ or \ha) \citep{Lampon2020, Lampon2021a, Lampon2022} or by using theoretical arguments about the upper limit of the heating efficiency \citep{Lampon2022}. For most of the studied planets, a large H/He ratio, larger than 97/3, has been found. In particular, for the sub-Neptune \gju12, we derived a value of 98/2 \citep{Lampon2022}. This planet is, among those analyzed, the one closest in size to HD\,235088\,b; hence, given the lack of further information on the H/He ratio, we use the same value in this analysis.

We found that the \het\ distribution of this planet is very extended, much more than in previously studied planets, including the sub-Neptune \gju12. 
An example of the modelled transmission for the combined transit (the average over the different phases) is shown in Fig.\,\ref{model_absorption} for a thermospheric temperature of 3000\,K and a sub-stellar mass-loss rate of 2.4\,$\times\,10^{10}$\,\gs. 
Because of the weak surface gravity of this planet, the velocities of the outflowing gas resulting from the hydrodynamic model 
are very large even at low radii, for instance, in the range of 5--15\,km\,s$^{-1}$ for $r$\,=\,2--15\,$R_{\rm P}$, thus 
inducing a very prominent broadening (compare blue and orange curves in Fig.\,\ref{model_absorption}).
As described in Sect.\,\ref{tra_spec}, the absorption peak is shifted to blue wavelengths by $-$6.6\,km\,s$^{-1}$, suggesting that a large fraction of the observed atmosphere is flowing towards the observer. Similar blue shifts have been found for most of the planets with \het\ detections. Since our 1D spherical and homogeneous model cannot predict it, we imposed a net shift of $-$6.6\,km\,s$^{-1}$ in our calculations.


With the method described above we constrain the mass-loss rate and temperature of the planet's upper atmosphere to the ranges of (1.5--5)\,$\times\,10^{10}$\,\gs\ and $T$\,=\,2750\,K--3500\,K (see Fig.\,\ref{chi2}). 
As this planet has a very extended atmosphere, the \het\ absorption at high altitudes is significant and, hence, 
it is advisable to estimate the potential effects of the stellar wind \citep[see e.g.][]{Vidotto2020,Lampon2022}. We did this by assuming that the atmosphere is still spherical but extended only up to the ionopause in the substellar direction \citep[see more details in][]{Lampon2022}. 
The results (displayed as the black triangles in Fig.\,\ref{chi2}) show that a strong SW does not significantly change our nominal \mlr-$T$ ranges.  
Furthermore, as we could not constrain the H/He ratio, we also explored the effects of varying this ratio (see Fig.\,\ref{chi2}).
The results show that increasing the H/He ratio to 99/1 does not significantly change the mass-loss rate. However, if the ratio were as low as 90/10 (unlikely given the previous results), then the mass-loss rate would be significantly lower.

\cite{Zhang_young_planets} estimated a temperature of 6700\,$\pm$\,300\,K and a mass-loss rate of $\sim$1.3\,$\times\,10^{11}$\,\gs\ for this planet. We derive significantly smaller values for both. Although we assumed different $F_{\rm XUV}$ stellar fluxes (by a factor of three; see above), we found that this does not explain the differences.
Instead, if we assume an upper boundary at 10–15\,\rp\ in our model and an H/He\,=\,90/10 value (\citealp{Zhang_young_planets} used 11\,\rp\ and H/He\,=\,90/10; M. Zhang, priv. comm. 2023) our results agree well. Nevertheless, as we would need an extraordinarily strong SW for confining the planetary wind at those altitudes, while current findings suggest a high H/He ratio, we think that this scenario is less likely and that the mass-loss rates and temperatures we derive are more plausible.

From a theoretical point of view, it is important to determine the hydrodynamic escape regime of the planets. 
Following the method used in previous studies \citep{Lampon2021b,Lampon2022}, 
we found that for the assumed H/He ratio of 98/2, this planet is in the photon-limited regime, with a heating efficiency of 0.23\,$\pm$\,0.03. However, for an H/He ratio of 90/10, although still in the photon-limited regime it approaches the energy-limited case, particularly at lower temperatures (3000\,K).



\section{Conclusions}

We derived new stellar parameters for the K-type star \hd23 and we obtained a new age estimation of 600--800\,Myr. Furthermore, by using multiband photometry from MuSCAT2, we confirmed the planetary nature of \hd23\,b and refined its planetary parameters. With a radius of $R_{\mathrm{p}}$\,=\,2.045$\pm$0.075\,R$_\oplus$, \hd23\,b is a young sub-Neptune planet close to the ``radius gap'' valley (\citealp{Fulton_2017, Fulton_2018}). More interesting is the confirmation with CARMENES spectra of an evaporating \ion{He}{I} atmosphere. Our excess absorption of $-$0.91$\pm$0.11$\%$ is $\sim$2$\sigma$ deeper than the previous detection with Keck/NIRSPEC by \citet{Zhang_young_planets}. The difference in the absorption depths suggests possible \ion{He}{I} variability, which should be clarified by further \ion{He}{I} observations.

We also analyzed the \ion{He}{I} signal detected in the transmission spectrum via hydrodynamical modeling. In comparison with previously studied planets \citep{Lampon2022}, the mass-loss rate and temperature of HD\,235088\,b are generally in the expected ranges. It has a low mass-loss rate corresponding to its moderate XUV irradiation level, and a low temperature as expected given its low gravitational potential. Its mass-loss rate and temperature are slightly lower than those of the sub-Neptunes \gj34 and \gju12.

There are only three exoplanets smaller than \hd23\,b with atmospheric detections\footnote{According to the ExoAtmospheres database (\url{http://research.iac.es/proyecto/exoatmospheres/index.php})}, and all are only tentative and are based on \textit{Hubble Space Telescope} or \textit{James Webb Space Telescope} observations: GJ\,1132\,b (1.16$\pm$0.11\,${R}_\oplus$, \citealp{GJ1132_HST}), GJ\,486\,b (1.34\,$\pm$\,0.06\,${R}_\oplus$, \citealp{GJ486b_JWST}), and LHS\,1140\,b (1.727$\pm$0.032\,${R}_\oplus$, \citealp{LHS1140b_HST}). 
In this work, we confirmed the presence of \ion{He}{I} in the atmosphere of \hd23\,b, making this planet the smallest one with such a robust atmospheric detection -- not only of \ion{He}{I} but of any atom or molecule.


\begin{acknowledgements}

CARMENES is an instrument at the Centro Astronómico Hispano en Andaluc\'\i{}a (CAHA) at Calar Alto (Almer\'{\i}a, Spain), operated jointly by the Junta de Andaluc\'ia and the Instituto de Astrof\'isica de Andaluc\'ia (CSIC).

CARMENES was funded by the Max-Planck-Gesellschaft (MPG), the Consejo Superior de Investigaciones Cient\'{\i}ficas (CSIC), the Ministerio de Econom\'ia y Competitividad (MINECO) and the European Regional Development Fund (ERDF) through projects FICTS-2011-02, ICTS-2017-07-CAHA-4, and CAHA16-CE-3978,  and the members of the CARMENES Consortium
(Max-Planck-Institut f\"ur Astronomie,
Instituto de Astrof\'{\i}sica de Andaluc\'{\i}a,
Landessternwarte K\"onigstuhl,
Institut de Ci\`encies de l'Espai,
Institut f\"ur Astrophysik G\"ottingen,
Universidad Complutense de Madrid,
Th\"uringer Landessternwarte Tautenburg,
Instituto de Astrof\'{\i}sica de Canarias,
Hamburger Sternwarte,
Centro de Astrobiolog\'{\i}a and
Centro Astron\'omico Hispano-Alem\'an), 
with additional contributions by the MINECO, 
the Deutsche Forschungsgemeinschaft (DFG) through the Major Research Instrumentation Programme and Research Unit FOR2544 ``Blue Planets around Red Stars'', 
the Klaus Tschira Stiftung, 
the states of Baden-W\"urttemberg and Niedersachsen, 
and by the Junta de Andaluc\'{\i}a.
We acknowledge financial support from the State Agency for Research of the Spanish MCIU (AEI) through projects
PID2019-110689RB-I00,
PID2019-109522GB-C51,
PID2019-109522GB-C5[4]/AEI/10.13039/501100011033, 
and the Centre of Excellence ``Severo Ochoa'' award to the
Instituto de Astrof\'isica de Andaluc\'ia (CEX2021-001131-S).

The first author acknowledges the special support from Padrina Conxa, Padrina Merc\`e, Jeroni, and Merc\`e. This work has made use of resources from AstroVallAlbaida-Mallorca collaboration. J.O.M. gratefully acknowledge the inspiring discussions with Maite Mateu, Alejandro Almod\'ovar, and Guillem Llodr\`a, and the warm support from Yess.

This work is partly supported by JSPS KAKENHI Grant Numbers P17H04574, JP18H05439, JP21K13955, and JST CREST Grant Number JPMJCR1761.
This paper is based on observations made with the MuSCAT2 instrument, developed by ABC, at Telescopio Carlos S\'anchez operated on the island of Tenerife by the IAC in the Spanish Observatorio del Teide.

This research was supported by the Excellence Cluster ORIGINS which is funded by the Deutsche Forschungsgemeinschaft (DFG, German Research Foundation) under Germany's Excellence Strategy - EXC-2094 - 390783311.
SC acknowledges support from DFG through project CZ 222/5-1. 

\end{acknowledgements}


\bibliographystyle{aa}
\bibliography{references}


\begin{appendix}
\label{Sec:Appendix}

\section{ Testing the MUV flux level }
\label{App: MUV flux}

\begin{figure}[ht!]
\includegraphics[width=1\linewidth]{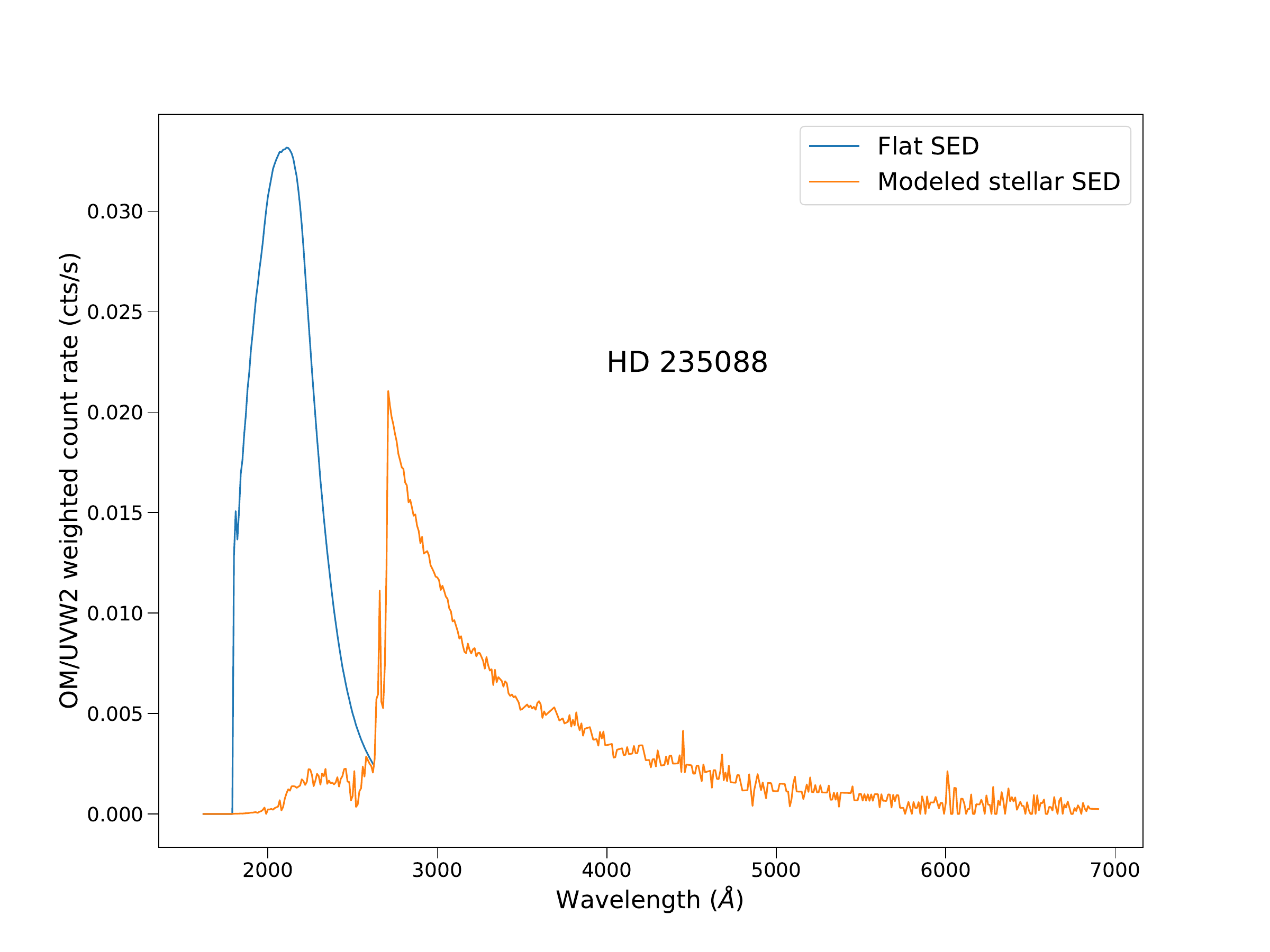}
\caption{XMM-Newton OM/UVW2 weighted count rate along the actual bandpass of the filter. The effective area is convolved with a flat SED emission (only in nominal band-pass) and with a realistic stellar emission (modelled as a black-body emission, at the effective temperature of the star, at $\lambda$\,$>$\,2700\,\AA). Most of the photons come from wavelengths outside the nominal band-pass ($\lambda \lambda$ 2120\,$\pm$\,500\,\AA). \label{fig:omseds}}
\end{figure}

One possibility for testing the actual level of the MUV flux would be the use of the UV filters onboard XMM-Newton, as suggested by \cite{Zhang_young_planets} and references therein. The XMM-Newton pipeline reports a flux calibrated assuming a flat SED in the whole spectral range of the filter (AB flux). For sources with a steep SED in the UV region, as in the case of late-type stars, this is not a valid approach. Instead, the effective area combined of the XMM-Newton Optical Monitor (OM) filter must be convolved with the stellar SED. A similar procedure was followed by \cite{Zhang_young_planets}. The XMM-Newton pipeline provides, for the source at HD\,235088 position, a count rate of 1.6\,cts/s in the UVW2 ($\lambda \lambda$ 2120\,$\pm$\,500\,\AA) filter, and 2.7\,cts/s in the UVM2 filter ($\lambda \lambda$ 2310\,$\pm$\,480\,\AA). Although the effective area of these filters is mainly within the nominal limits, there is a non-negligible tail towards longer wavelengths.

To test the wavelengths  at which the photons recorded in the OM/UVW2 filter had originated in, we extended our modelled SED up to 7000\,\AA,\ assuming a black-body emission for a star with temperature and size values as listed in Table\,\ref{table - System Param}. We applied a 30\% reduction of the efficiency of the instrument due to degradation of the CCD\footnote{\url{https://xmmweb.esac.esa.int/docs/documents/CAL-SRN-0378-1-1.pdf}} over time. If we limit our test to the nominal UVW2 band-pass, we obtain 0.14\,cts/s, just a $\sim$9\% of the observed count rate, but the use of the SED in the whole 1600--7000\,\AA\ spectral range result in a count rate of 1.59\,cts/s.  Same procedure with the OM/UVM2 filter yields 2.7\,cts/s. Both values are in excellent agreement with the observed count rate. This implies that no correction is needed to the general level of our SED.

Figure\,\ref{fig:omseds} shows the weighted count rate after convolving the effective area of the OM/UVW2 filter and an assumed SED. We used a flat SED (useful for AB magnitude or flux) in the nominal band-pass, and a realistic stellar SED in the whole spectral range. Although both combinations result in the same accumulated count rate, they yield quite different fluxes given the different distribution of photons along the spectrum. We find that 90\% of the photons are actually coming from a wavelength range outside the nominal band-pass. This test indicates that the use of these filters to evaluate the stellar flux in the UV band must be taken with care.

\section{ Photometric fit extra figures}
\label{App: TOI-1430 juliet}

\begin{figure*}[ht!]
     \centering
     \begin{subfigure}{0.32\textwidth}
         \centering
         \includegraphics[width=\textwidth]{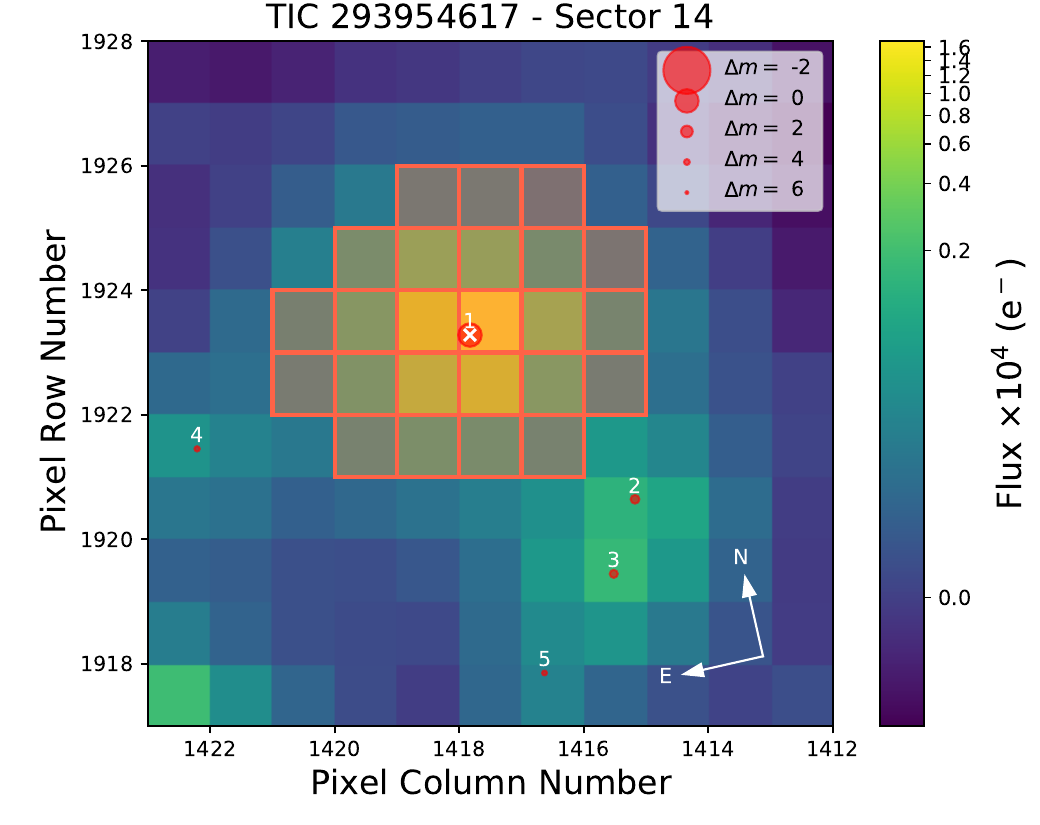}
     \end{subfigure}
     \hfill
     \begin{subfigure}{0.32\textwidth}
         \centering
         \includegraphics[width=\textwidth]{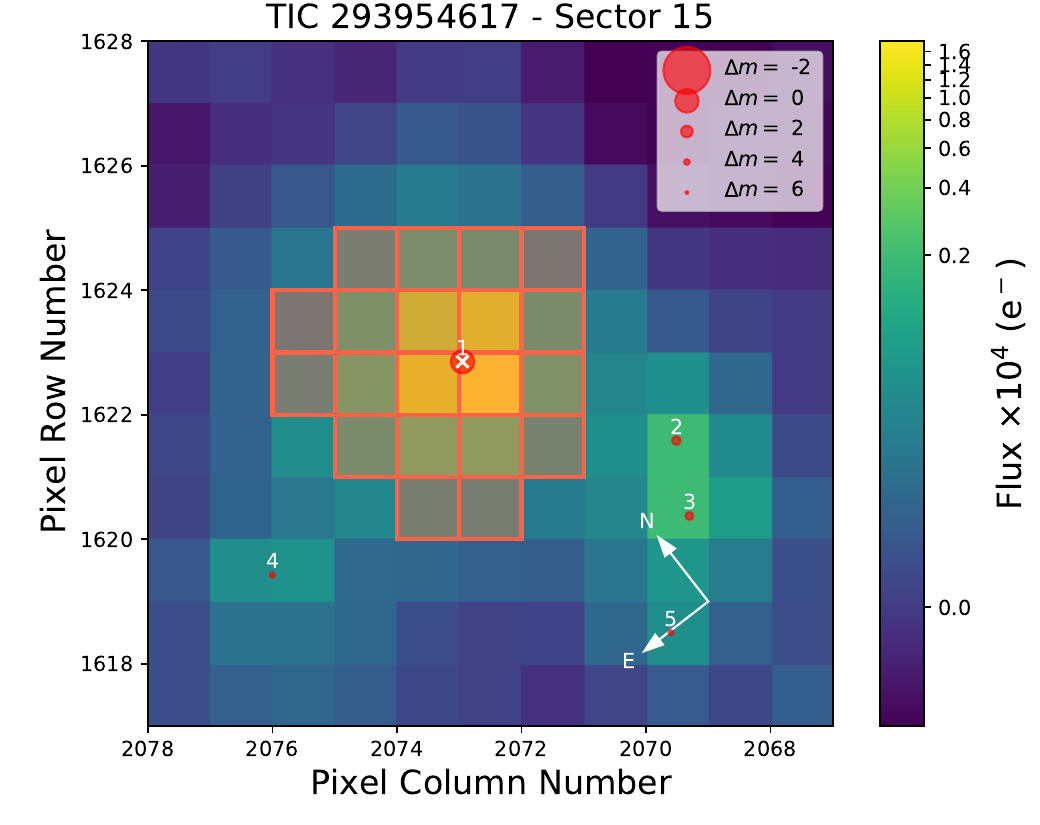}
     \end{subfigure}
     \hfill
     \begin{subfigure}{0.32\textwidth}
         \centering
         \includegraphics[width=\textwidth]{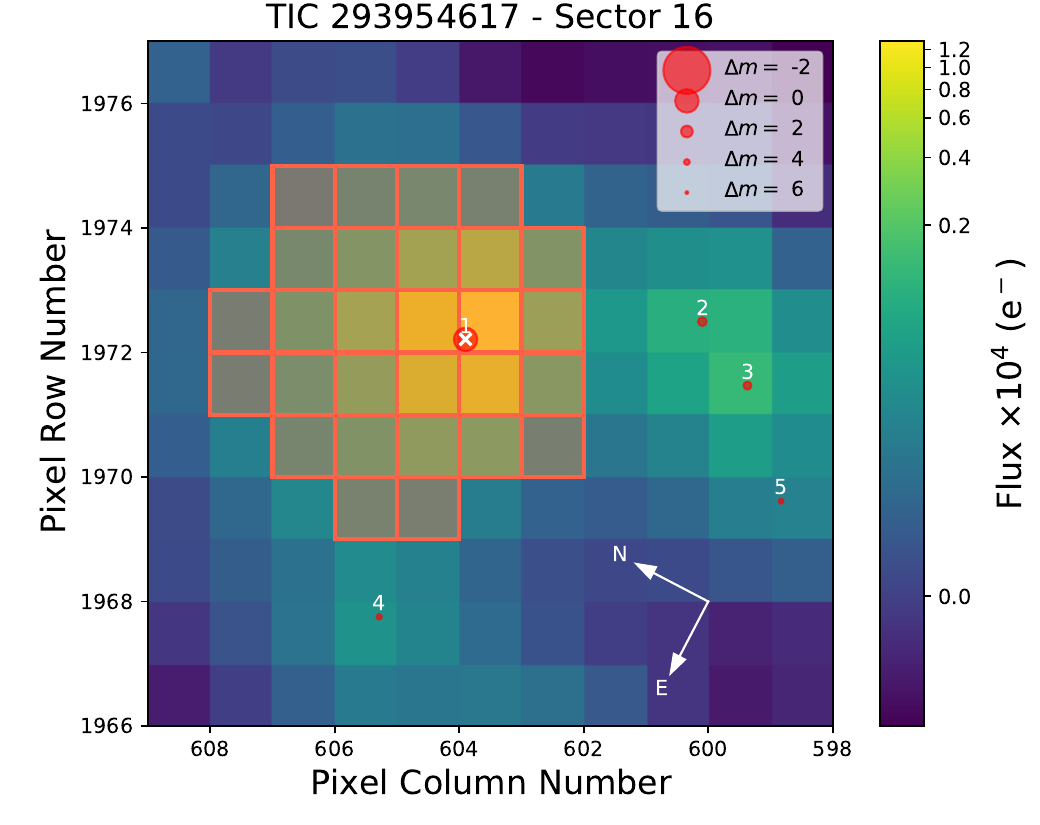}
     \end{subfigure}
     \hfill
     \begin{subfigure}{0.32\textwidth}
         \centering
         \includegraphics[width=\textwidth]{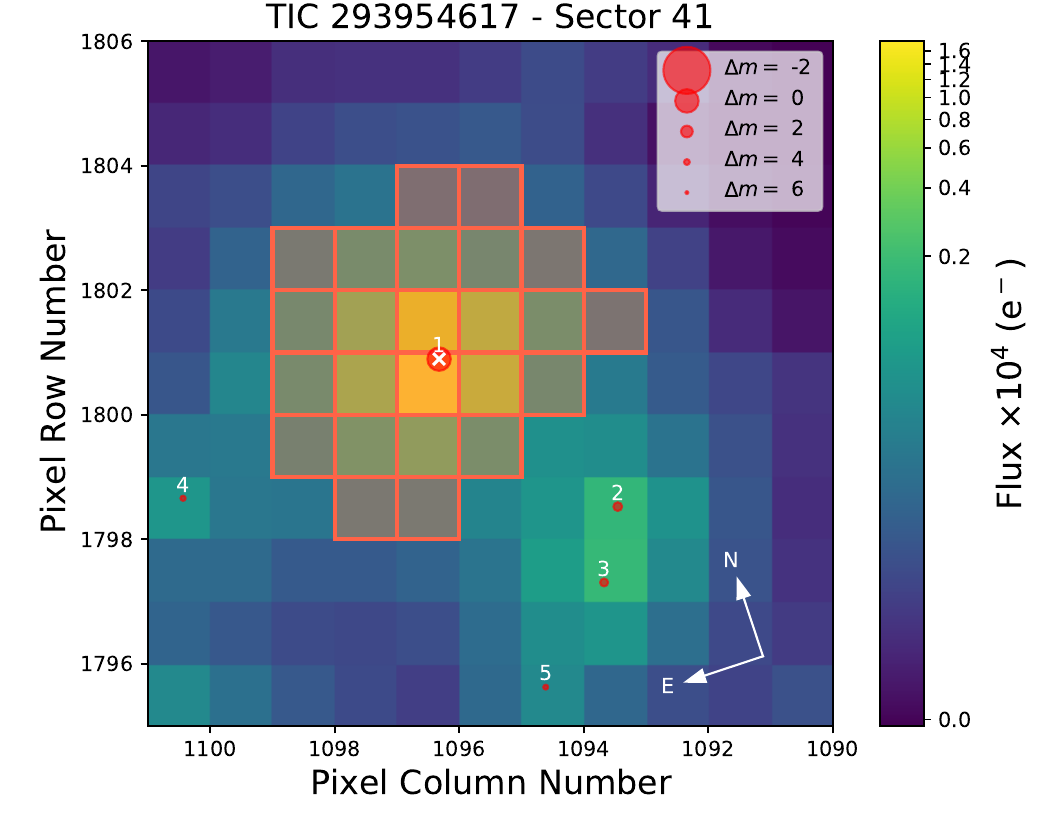}
     \end{subfigure}
     \hfill
     \begin{subfigure}{0.32\textwidth}
         \centering
         \includegraphics[width=\textwidth]{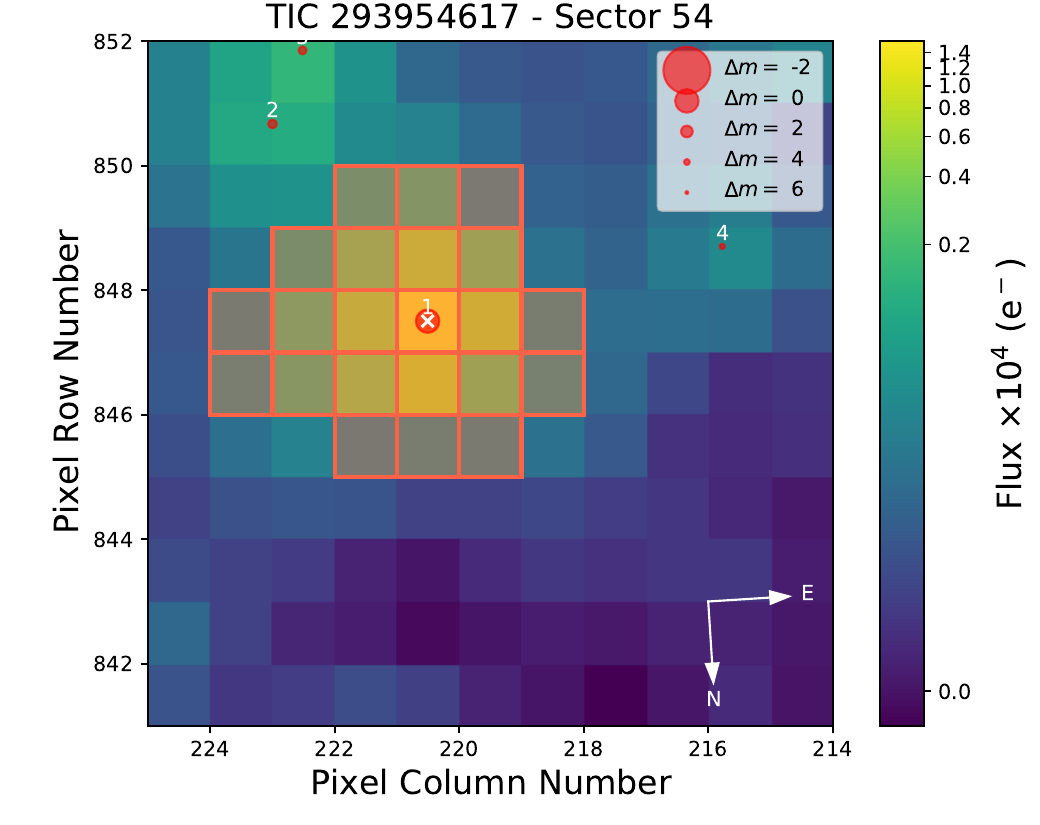}
     \end{subfigure}
     \begin{subfigure}{0.32\textwidth}
         \centering
         \includegraphics[width=\textwidth]{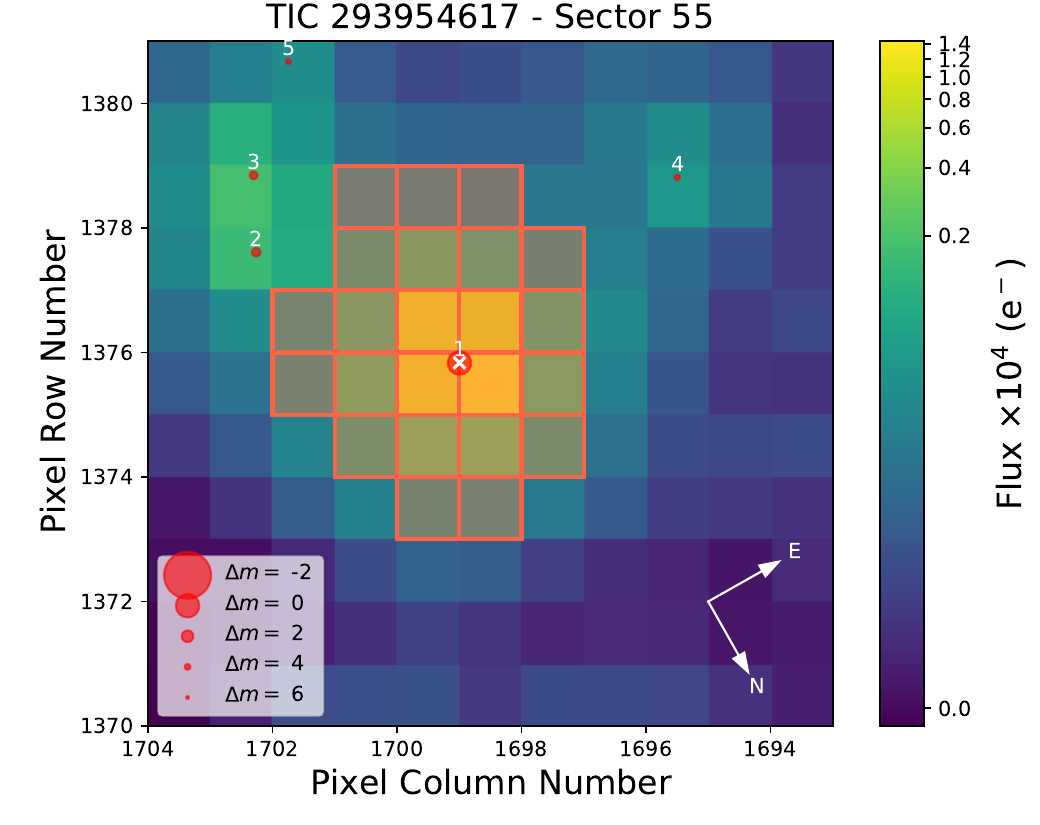}
     \end{subfigure}
     \begin{subfigure}{0.32\textwidth}
         \centering
         \includegraphics[width=\textwidth]{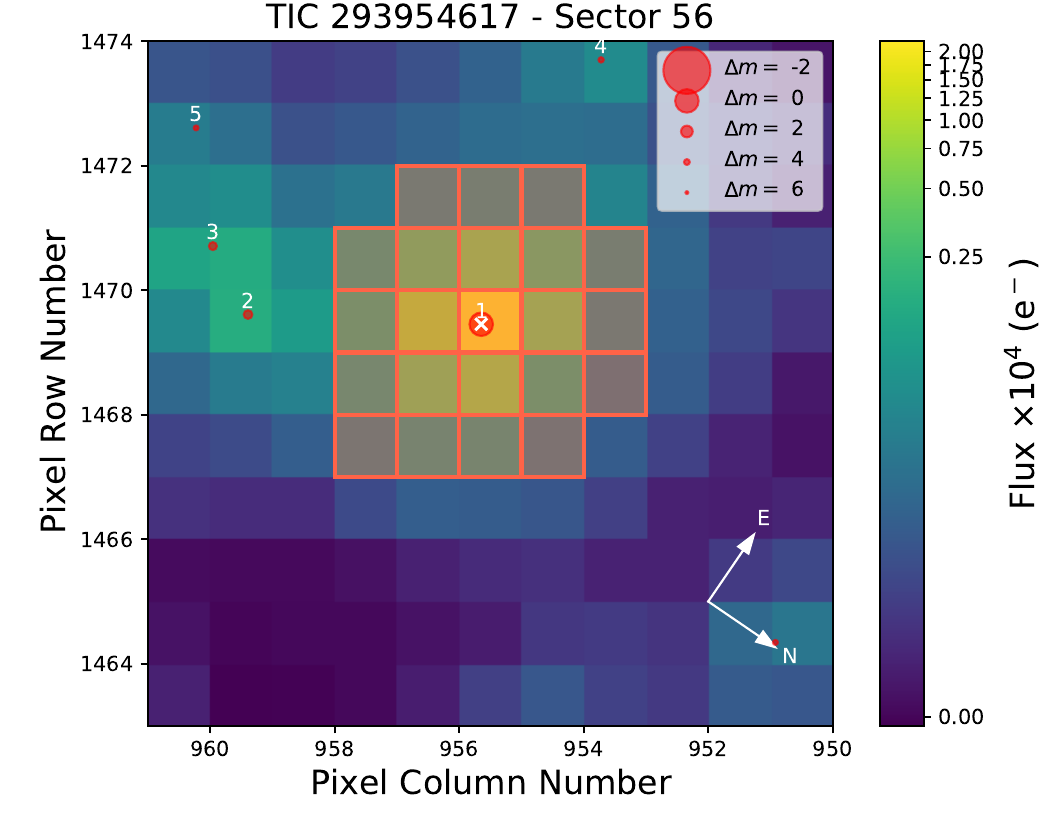}
     \end{subfigure}
    \caption{\textit{TESS} target pixel file image of \hd23 (TIC\,293954617) observed in Sectors\,14, 15, 16, 41, 54, 55, and 56 (made with \texttt{tpfplotter}). The pixels highlighted in red show the aperture used by \textit{TESS} to get the photometry. The electron counts are color-coded. The position and sizes of the red circles represent the position and \textit{TESS} magnitudes of nearby stars respectively. \hd23 is marked with a '$\times$' and labeled as $\#$1.}
        \label{Fig: TESS_FOV}
\end{figure*}

\begin{figure*}[ht!]
    \centering
    \includegraphics[width=1\linewidth]{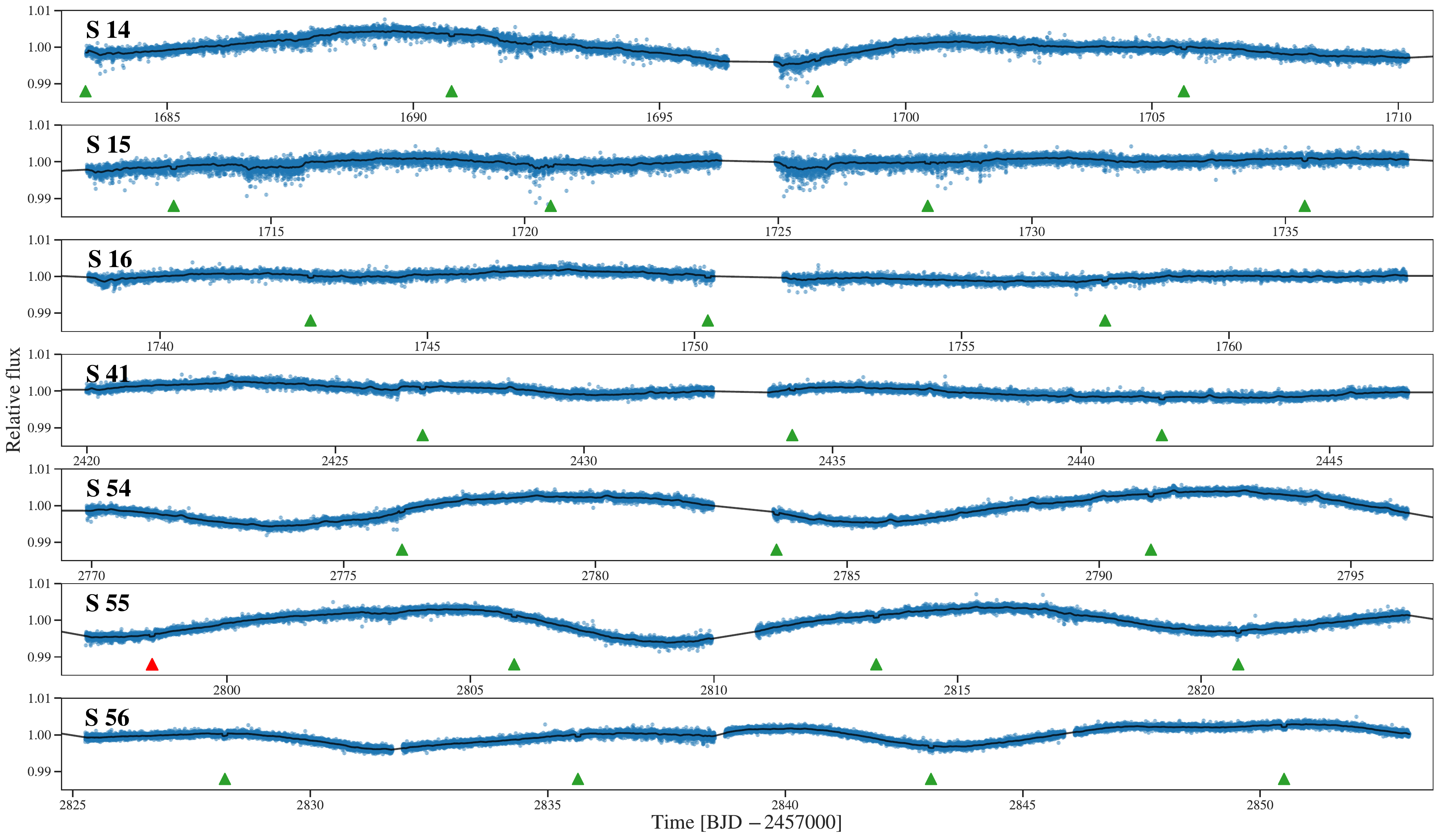}
    \caption{\label{Fig: TOI-1430 TESS fit}
    \hd23 two-min cadence SAP \textit{TESS} photometry from Sectors\,14, 15, 16, 41, 54, 55, and 56 along with the transit plus GP model. Upward-pointing green triangles mark \hd23\,b's transits. Upward-pointing red triangle marks the transit observed with CARMENES, and analyzed in this work.
    }
\end{figure*}


\section{Multi-color validation}
\label{App: multicolor_validation}

\begin{table}[ht!]
\footnotesize
\caption[width=\textwidth]{
\label{Table: priors_posteriors} Prior and posterior distributions from the \texttt{eemce} algorithm for the \hd23\,b.}

\centering

\begin{tabular}{lcc}

\hline \hline 
\noalign{\smallskip} 

Parameter & Prior & Posterior \vspace{0.05cm}\\
\hline
\noalign{\smallskip}

$P$ ~[d] & $\mathcal{N}(7.434162, 4.12\times10^{-5})$ & 7.434128\,(6) \vspace{0.05cm}\\
$b$ & $\mathcal{U}(0,1)$ & $0.30^{+0.25}_{-0.21}$ \vspace{0.05cm} \\ 
$t_0$ ~[BJD] & $\mathcal{N}(2459850.501,0.003)$ & 2459850.502\,(1) \vspace{0.05cm} \\
$T_{\rm eff}$ & $\mathcal{N}(5064,119)$ & $5040 \pm 120$ \vspace{0.05cm}\\

\noalign{\smallskip} 
\hline 
\noalign{\smallskip} 
\multicolumn{3}{c}{\textit{ Derived planetary parameters }} \\ 
\noalign{\smallskip}

$a/{R}_{\star}$ & -- & $20.90^{+0.90}_{-2.50}$ \vspace{0.05cm}\\
$a_{\rm p}$ ~[AU] & -- & $0.0757^{+0.0032}_{-0.0091}$ \vspace{0.05cm}\\
$i_{\rm p}$ ~(deg) & -- & $89.20^{+0.50}_{-0.90}$ \vspace{0.05cm}\\

\noalign{\smallskip}
\hline
\end{tabular} 
\tablefoot{ Prior labels $\mathcal{U}$ and $\mathcal{N}$ represent the uniform and normal distribution, respectively }
\end{table}

\begin{figure*}
    \centering
    \includegraphics[width=\hsize]{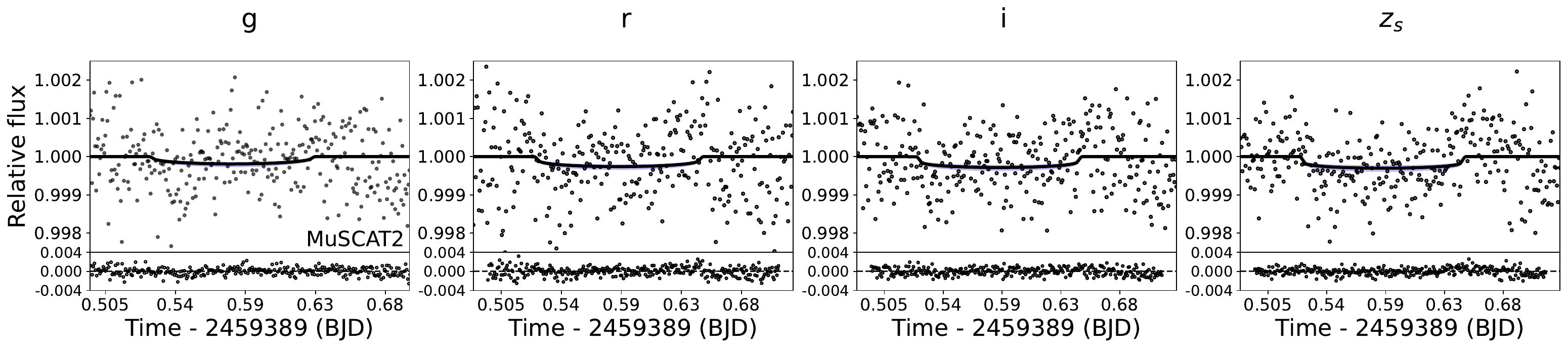}
    \caption{\label{fig:my_label} MuSCAT2 unfolded light-curves on each filter \textit{g}, \textit{r}, \textit{i} and \textit{$z_s$} for the \hd23\,b. We have plotted $1\sigma$ uncertainties for the transit on each passband.}
\end{figure*}


\section{Additional figures and tables}

\begin{figure}
    \centering
   \begin{subfigure}{0.49\textwidth}
         \centering
         \includegraphics[width=\textwidth]{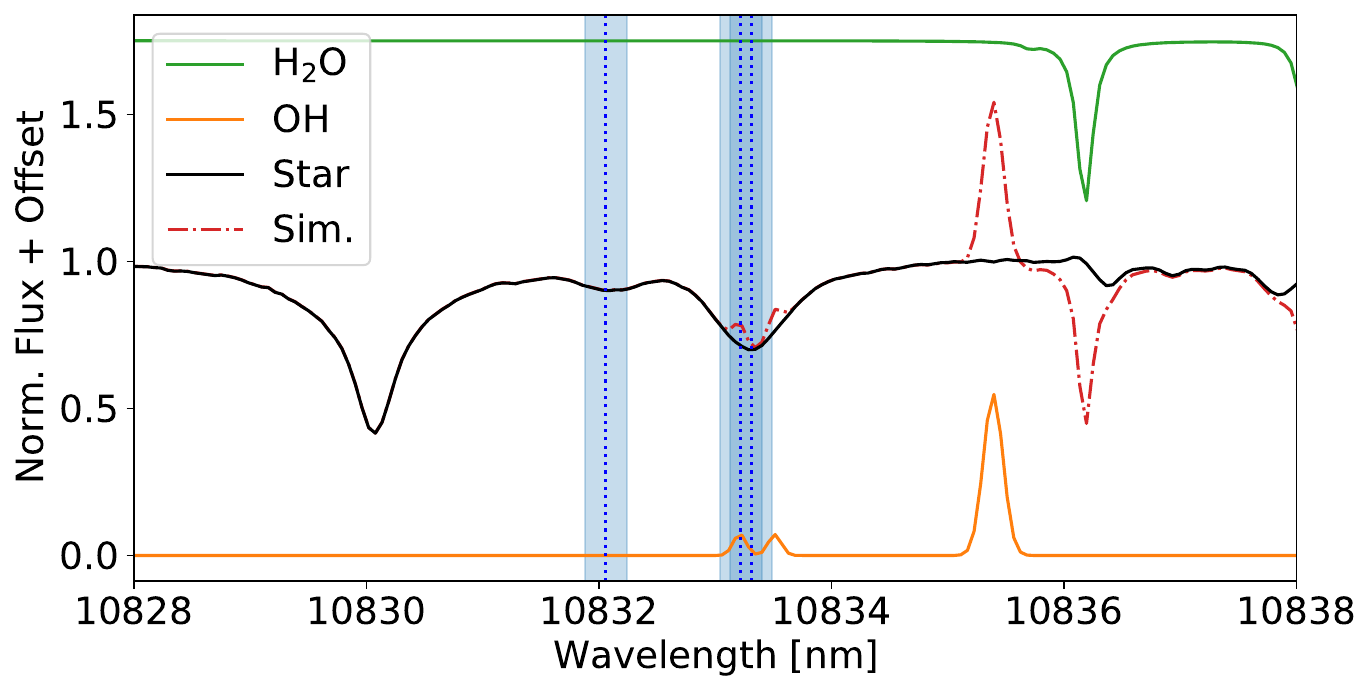}
     \end{subfigure}
     \hfill
     \begin{subfigure}{0.49\textwidth}
         \centering
         \includegraphics[width=\textwidth]{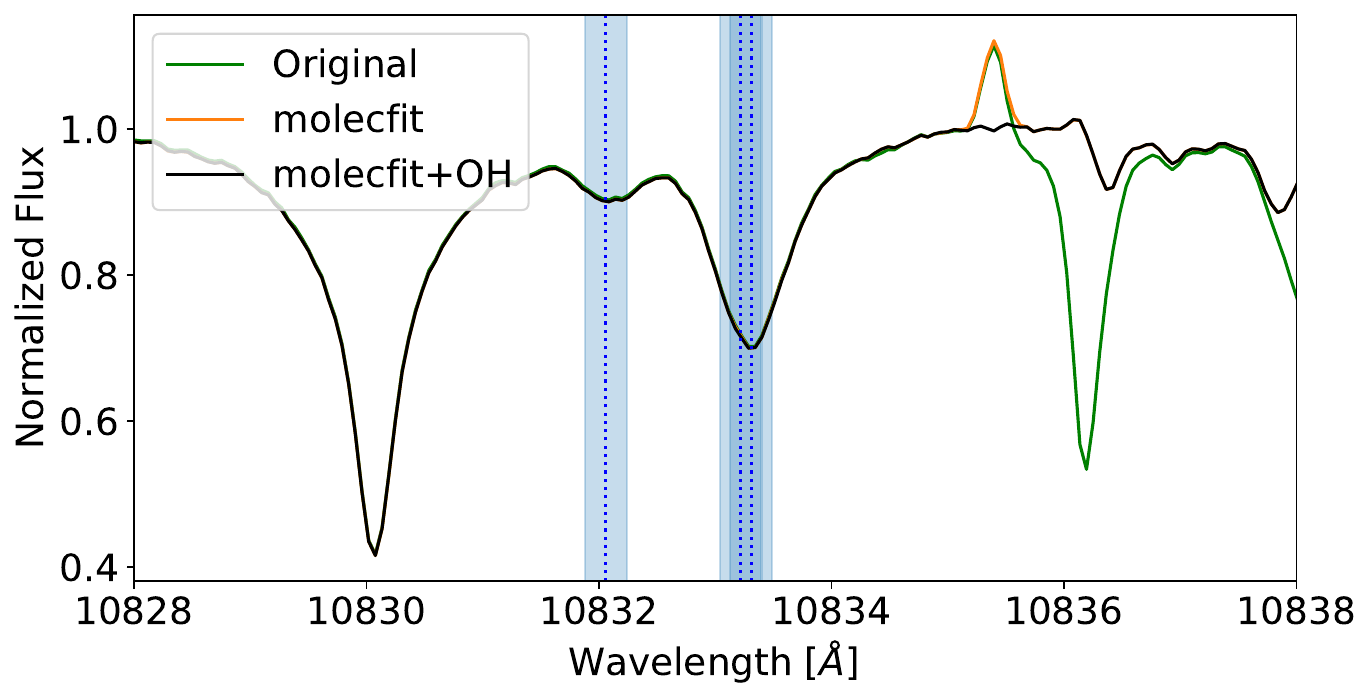}
     \end{subfigure}
    \caption{Telluric contamination close to the \ion{He}{I} triplet lines. \textit{Upper panel:} Simulation of the contamination of the spectrum of \hd23 by H$_2$O absorption and OH emission during the night of 6 August 2022. The green curve is a synthetic model of H$_2$O absorption, the orange curve is a synthetic model of OH emission, and the black curve is the average of the normalised \hd23 spectra. The dashed red line is the combination of synthetic telluric models and the \hd23 spectrum.
    \textit{Lower panel:} Averaged normalised \hd23 spectra from 6 August 2022 CARMENES observations. Original spectrum is plotted in green, spectrum after \texttt{molecfit} correction is over-plotted in orange, and spectrum after \texttt{molecfit} and OH correction is over-plotted in black.
    The vertical blue dotted lines indicate the positions of the \ion{He}{I} triplet lines, and the blue shaded region represents the planet trace in the stellar rest frame at vacuum wavelength.}
    \label{Fig: telluric simulation}
\end{figure}

\begin{figure}
\includegraphics[width=1\linewidth]{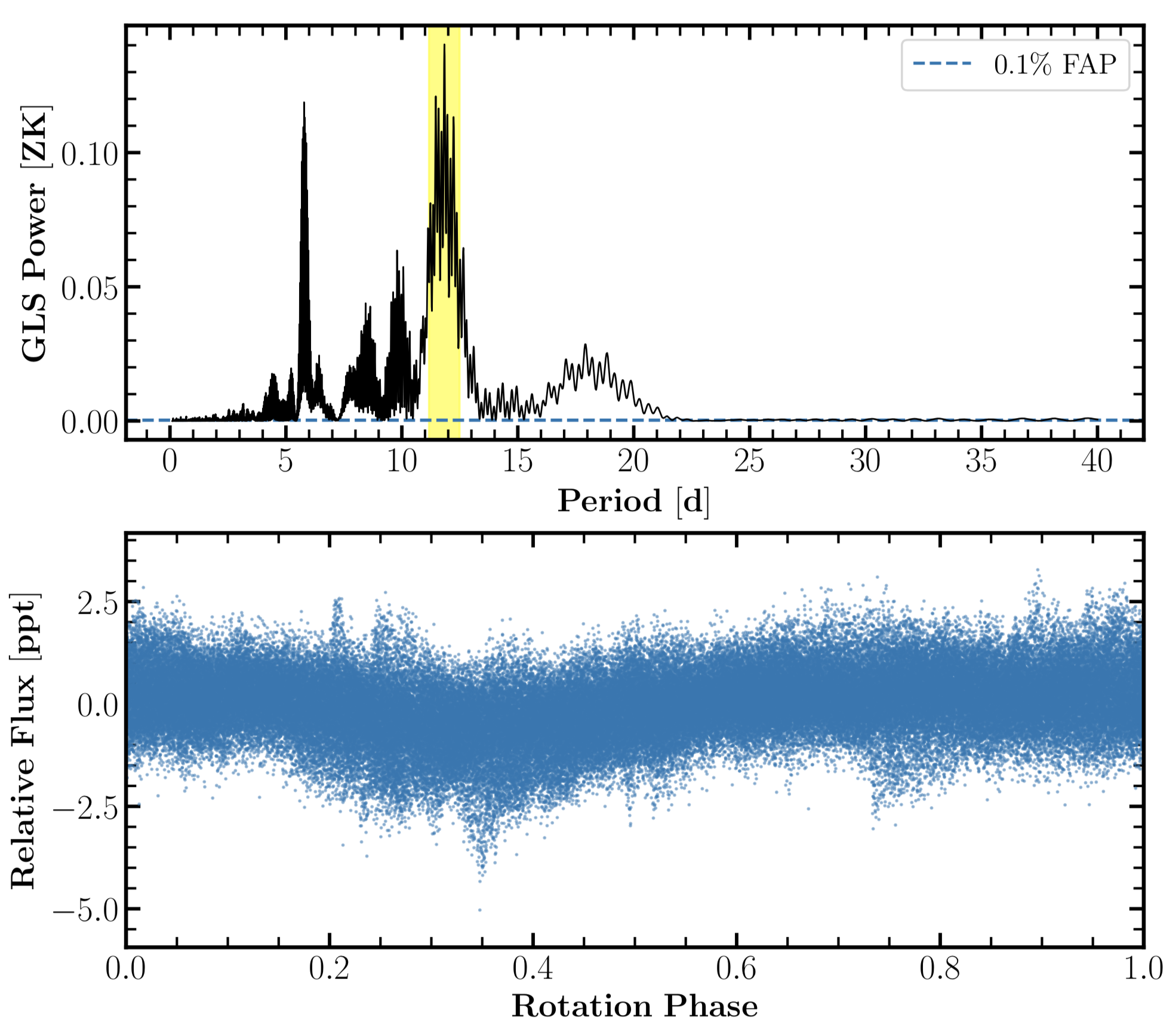}
\caption{ GLS periodogram of the \textit{TESS} photometry that includes the combination of all sectors (\textit{top panel}). The yellow vertical band shows the maximum peak found, whose period we associate with the rotation period of the star. Phase-folded plot for the period of 11.8\,days (\textit{bottom panel}). \label{fig:GLS_Prot}}
\end{figure}

\begin{table}[ht!]
\caption[width=\textwidth]{
\label{table - TOI-1430 priors and posteriors}
Prior and posterior distributions from the nested sampling fitting for \hd23\,b \ion{He}{I} signal (see Fig.\,\ref{Fig: TS PLOTS}).}
\centering

\begin{tabular}{lcc}

\hline \hline 
\noalign{\smallskip} 

Parameter & Prior & Posterior \vspace{0.05cm}\\
\hline
\noalign{\smallskip}

Absorption ~[\%] & $\mathcal{U}(-3, 3)$ & $-$0.91$^{+0.10}_{-0.11}$  \vspace{0.05cm} \\ 
$\lambda_0$ ~[\AA] & $\mathcal{U}(10830,10835)$ & 10832.98$\pm$0.05 \vspace{0.05cm} \\ 
$\sigma$ ~[\AA] &  $\mathcal{U}(0.0,1)$ & 0.39$^{+0.07}_{-0.06}$  \vspace{0.05cm} \\ 

$\Delta$$v$ ~[km\,s$^{-1}$] & -- & $-$6.6$\pm$1.3 \\
FWHM ~[\AA] & -- &  0.95$^{+0.16}_{-0.14}$ \\
EW ~[m\AA] &  -- & 9.5$^{+1.1}_{-1.0}$  \vspace{0.05cm} \\ 

\noalign{\smallskip}
\hline
\end{tabular}

\tablefoot{ Prior label $\mathcal{U}$ represents uniform distribution. }
\end{table}

\begin{figure}
    \centering
    \includegraphics[width=\hsize]{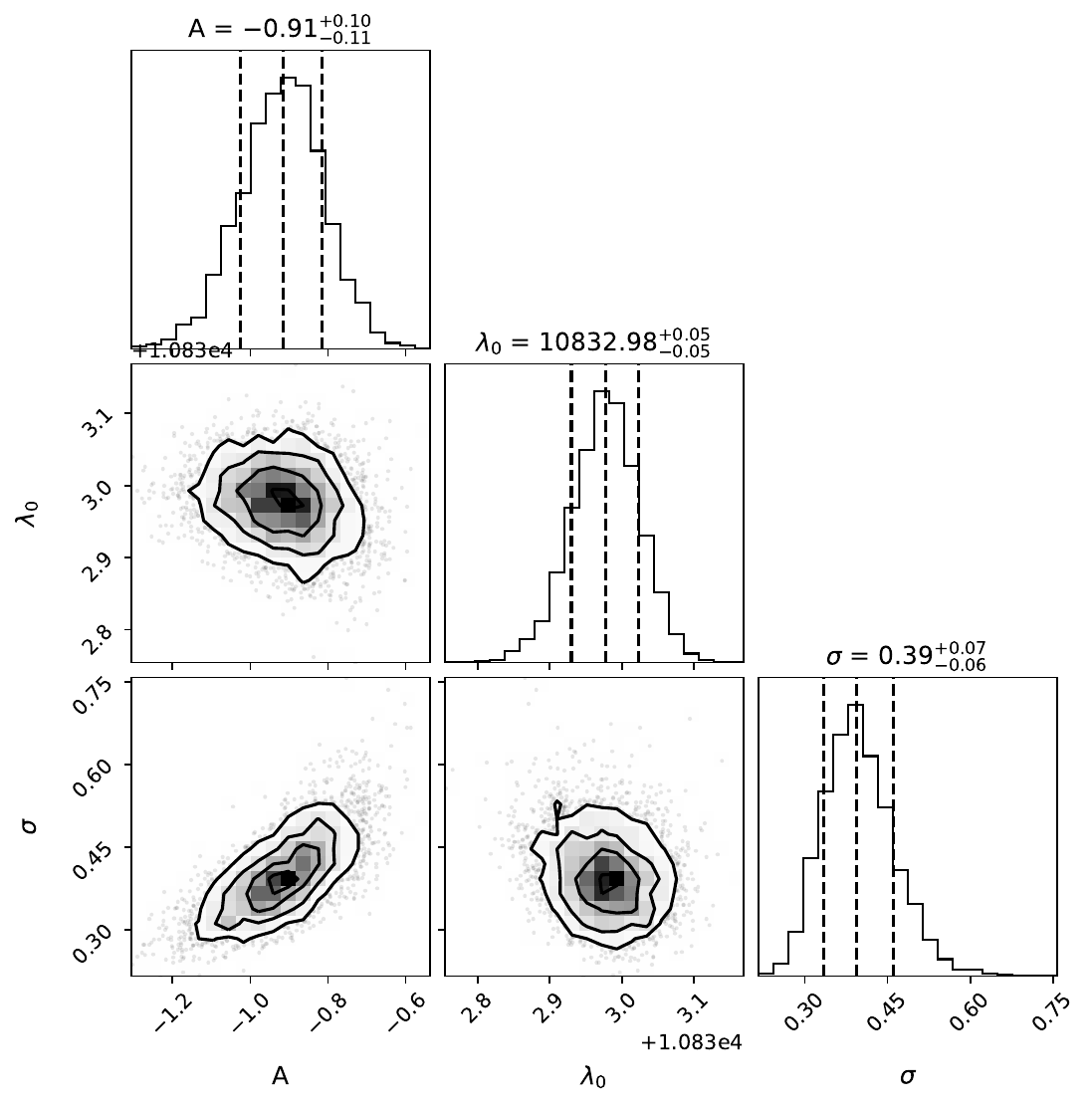}
    \caption{\label{Fig: TOI-1430 nested Helium}
    Corner plot for the nested sampling posterior distribution of the \hd23\,b \ion{He}{I} signal.
    }
\end{figure}

\begin{figure}
    \centering
    \includegraphics[width=\hsize]{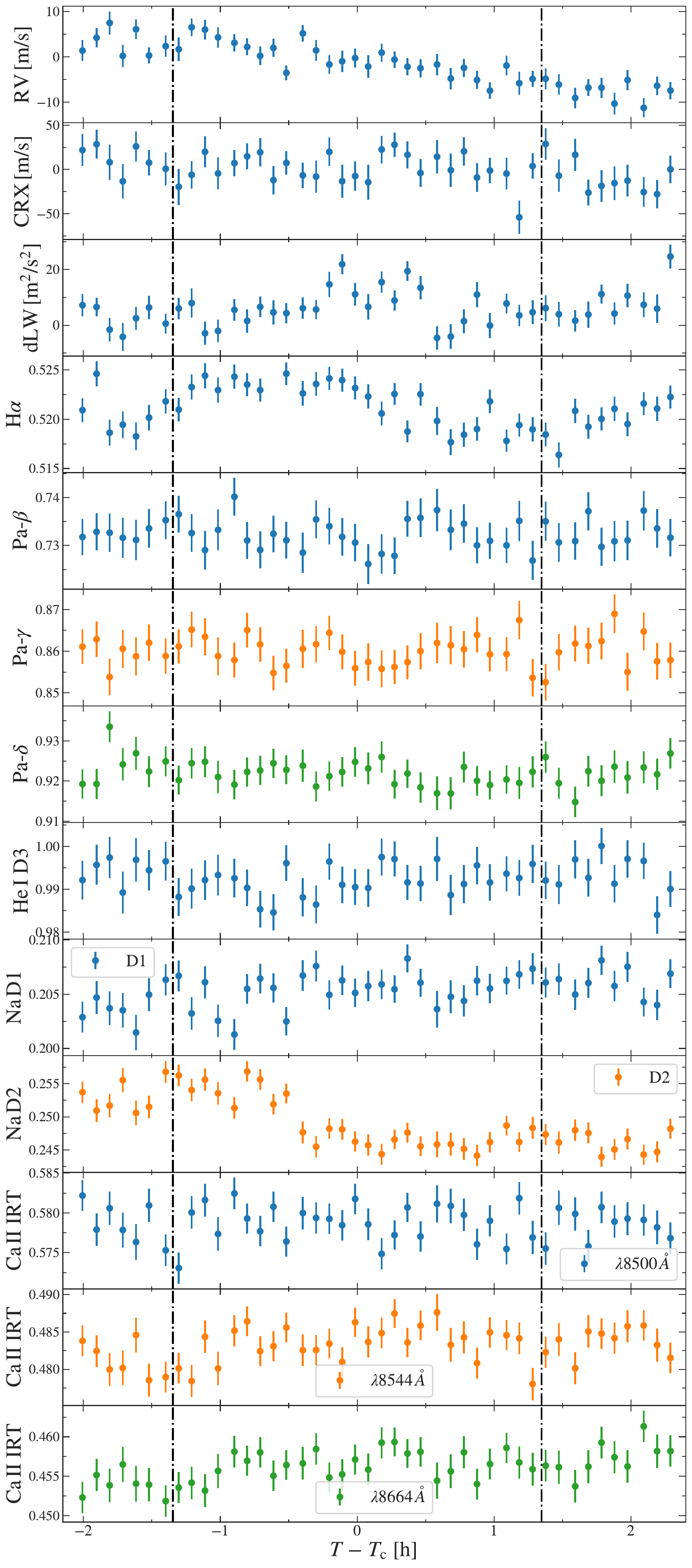}
    \caption{\label{Fig: Activity}
    Time evolution of the \texttt{serval} products: RVs from VIS, activity indicators, and line indices; and the light curves of the \ion{H}{}\,Paschen\,$\beta$ (Pa-$\beta$, 12821.6\,\AA; fifth panel), \ion{H}{}\,Paschen\,$\gamma$ (Pa-$\gamma$, 10941.1\,\AA; sixth panel), and \ion{H}{}\,Paschen\,$\delta$ (Pa-$\delta$, 10052.1\,\AA; seventh panel) lines, and the \ion{He}{I}\,D3 line at 5877.2\,$\AA$ (eighth panel). Vertical dash dotted black lines indicate the first and fourth contacts.
    }
\end{figure}

\end{appendix}

\end{document}